\documentclass[twocolumn]{aastex631}

\newcommand{\htwo}{\mbox{H$_2$}}

\newcommand{\cii}{\mbox{[C{\small II}]}}

\newcommand{\hii}{\mbox{H{\small II}}}
\newcommand{\heii}{\mbox{He{\small II}}}

\newcommand{\msun}{\mbox{\,M$_{\odot}$}}
\newcommand{\lsun}{\mbox{\,$\rm L_{\odot}$}}

\newcommand{\kms}{km\,s$^{-1}$}

\newcommand{\spitzer}{\textit{Spitzer}}
\newcommand{\herschel}{\textit{Herschel}}
\newcommand{\cm}{\mbox{$\mathrm{cm^{-3}}$}}

\newcommand\changes{}
\newcommand\change{}

\newcommand\newchanges{}

\usepackage{enumitem}
\usepackage{appendix}
\usepackage{hyperref}
\usepackage{amsmath,amstext}

\shorttitle{}
\shortauthors{}
\graphicspath{{./}{figures/}}
\accepted{April 10, 2024}
\submitjournal{ApJ}

\shortauthors{Tarantino et al.}

\begin{document}

\title{Modeling Ionized Gas in the Small Magellanic Cloud: The Wolf-Rayet Nebula N76}

\author[0000-0003-1356-1096]{Elizabeth Tarantino}
\affiliation{Space Telescope Science Institute, 3700 San Martin Drive, Baltimore, MD 21218, USA}
\affiliation{Department of Astronomy, University of Maryland, College Park, MD 20742, USA}

\author[0000-0002-5480-5686]{Alberto D. Bolatto}
\affiliation{Department of Astronomy, University of Maryland, College Park, MD 20742, USA}

\author[0000-0002-4663-6827]{R\'{e}my Indebetouw}
\affiliation{Department of Astronomy, University of Virginia, Charlottesville, VA 22904, USA}
\affiliation{National Radio Astronomy Observatory, 520 Edgemont Road, Charlottesville, VA 22903, USA}

\author[0000-0002-5307-5941]{M\'{o}nica Rubio}
\affiliation{Departamento de Astronom\'{i}a, Universidad de Chile, Casilla 36-D, Santiago, Chile}

\author[0000-0002-4378-8534]{Karin M. Sandstrom}
\affiliation{Center for Astrophysics \& Space Sciences, Department of Physics, University of California, San Diego, 9500 Gilman Drive, San Diego, CA 92093, USA}

\author[0000-0003-1545-5078]{J.-D T. Smith}
\affiliation{Dept. of Physics \& Astronomy, University of Toledo, Toledo, OH 43606, USA}

\author[0009-0000-4156-5604]{Daniel Stapleton}
\affiliation{Space Telescope Science Institute, 3700 San Martin Drive, Baltimore, MD 21218, USA}
\affiliation{Department of Astronomy, University of Maryland, College Park, MD 20742, USA}

\author[0000-0002-5480-5686]{Mark Wolfire}
\affiliation{Department of Astronomy, University of Maryland, College Park, MD 20742, USA}

\correspondingauthor{Elizabeth Tarantino}
\email{etarantino@stsci.edu}

%% Mark off the abstract in the ``abstract'' environment. 
\begin{abstract}

% The main goal of this work is to model the physical conditions of the ionized gas in N76 to provide a detailed, resolved view of the impact the most energetic stars have on the low metallicity ISM. Most previous studies focused on understanding the ionized gas properties in low metallicity galaxies analyze unresolved spectra that contain emission from different phases of the ISM and a variety of HII regions that vary in properties blended into one beam. Further, the galaxies in these studies are typically undergoing a burst of star-formation and contain a large number of WR stars. The approach in this work is to model N76 to isolate the impact WR stars have on the low metallicity ISM, as it is the closest low metallicity Wolf-Rayet nebula that is powered by a single well characterized ionization source. 

We present {\tt Cloudy} modeling of infrared emission lines in the Wolf-Rayet (WR) nebula N76 caused by one of the most luminous \newchanges{and hottest} WR stars in the low metallicity Small Magellanic Cloud. We use spatially resolved mid-infrared Spitzer/IRS and far-infrared Herschel/PACS spectroscopy to establish the physical conditions of the ionized gas. The spatially resolved distribution of the emission allows us to constrain properties much more accurately than using spatially integrated quantities. We construct models with a range of constant \newchanges{hydrogen} densities between $\rm n_H = 4 - 10 \ \cm$ and a stellar wind-blown cavity of 10~pc which reproduces the intensity and shape of most ionized gas emission lines, including \changes{the high ionization lines [\ion{O}{4}] and [\ion{Ne}{5}], as well as [\ion{S}{3}], [\ion{S}{4}], [\ion{O}{3}], and [\ion{Ne}{3}].} Our models suggest that the majority of [\ion{Si}{2}] emission (91\%) is produced at the edge of the HII region \newchanges{around} the transition between ionized and atomic gas \newchanges{while very little of the [\ion{C}{2}] emission ($<5\%$) is associated with the ionized gas.} \newchanges{The physical conditions of N76 are characterized by a hot \hii\ region with a maximum electron temperature of $T_e \sim 24,000 \rm \, K$, electron densities that range from $n_{e} \sim 4 \ \rm to \ 12 \rm \, cm^{-3}$, and high ionization parameters of $\log(U) \sim -1.15  \ \rm to \ -1.77$. By analyzing a low metallicty WR nebula with a single ionization source, this work gives valuable insights on the impact WR stars have to the galaxy-integrated ionized gas properties in nearby dwarf galaxies. }

% These observations of an isolated, low metallicity WR nebula give valuable insight to interpreting galaxy integrated ionized gas properties of blue compact dwarf galaxies. 

% The physical conditions of N76 derived from the {\tt Cloudy} models characterize a hot \hii\ region with a maximum electron temperature range from $T_e \sim 22,000-24,000 \rm K$, a relatively small density range of $n_{e} \sim 4-12 \ \rm cm^{-3}$, and high ionization parameters of $U = 1.7 \times 10^{-2}$ to $U = 7.1 \times 10^{-2}$ 

% ionized gas in N76 makes a small ($<5\%$) contribution to the neutral-dominated lines, [\ion{C}{2}] and [\ion{O}{1}], while producing most (91\%) of the [\ion{Si}{2}] emission. 
%Overall, N76 is a lower density ($n_e \approx 10 \ \cm$), large (40~pc), WR nebula, but generally agrees with the ionized gas properties in other nearby low metallicity systems and has a similar structure to other WR nebula. 

%Previous studies of mostly unresolved infrared spectroscopy in low metallicity star-forming galaxies find harder radiation fields, extended bright [\ion{O}{3}] emission, and an overall more porous structure. 
%We model N76 as a spherically symmetric HII region with a WR and O6 supergiant star binary system named AB7 as the ionization source using the photoionization code .

\end{abstract}

%% Keywords should appear after the \end{abstract} command. 
%% The AAS Journals now uses Unified Astronomy Thesaurus concepts:
%% https://astrothesaurus.org
%% You will be asked to selected these concepts during the submission process
%% but this old "keyword" functionality is maintained in case authors want
%% to include these concepts in their preprints.
\keywords{Interstellar medium (847), H II regions (694), Small Magellanic Cloud (1468), Wolf-Rayet stars (1806), Photoionization (2060) }
%% From the front matter, we move on to the body of the paper.
%% Sections are demarcated by \section and \subsection, respectively.
%% Observe the use of the LaTeX \label
%% command after the \subsection to give a symbolic KEY to the
%% subsection for cross-referencing in a \ref command.
%% You can use LaTeX's \ref and \label commands to keep track of
%% cross-references to sections, equations, tables, and figures.
%% That way, if you change the order of any elements, LaTeX will
%% automatically renumber them.
%%
%% We recommend that authors also use the natbib \citep
%% and \citet commands to identify citations.  The citations are
%% tied to the reference list via symbolic KEYs. The KEY corresponds
%% to the KEY in the \bibitem in the reference list below. 

\section{Introduction} \label{sec:intro}

Massive stars inject energy into the surrounding gas, heating and influencing the chemistry of their nearby interstellar medium (ISM). The effects massive stars have on the ISM depend on a variety of factors, including the metallicity of the surrounding material. Lower metallicity stars have harder radiation fields and a decreased mass-loss rate \citep[e.g.,][]{Hurley2000, Vink2001}. At the same time, the low metallicity ISM contains less dust \citep[e.g.,][]{Galliano2021}, which consequently leads to a more porous structure \changes{with a lower HII region \newchanges{effective} covering fraction} allowing radiation fields to penetrate deeper into the ISM \citep[e.g.,][]{Poglitsch1995, Cormier2015, Cormier2019}. The chemistry and physical processes governing the interaction between massive stars is thus highly affected by the metallicity of the medium. 

% Further, nearby low metallicity environments allow us to study the ISM and we can apply this insight to low metallicity, high redshift galaxies. 
%  Understanding the combined affect of low metallicity systems is essential to gaining insight on low metallicity galaxies at high redshifts. while it is challenging to examine the ISM of the high red shift galaxies directly, local low metallicity galaxies can be examined in detail to determine the effect of metallicity on ISM properties.  

Investigation of ionized gas tracers are used to determine the properties of the ionized gas across galaxies. In particular, the infrared (IR) lines such as [\ion{Ne}{2}], [\ion{S}{3}], [\ion{O}{3}], [\ion{Ne}{3}], and [\ion{S}{4}], accessed by the \spitzer, \herschel, and \textit{James Webb} space telescopes trace a variety of conditions in the ISM, including the ionizing radiation field strength and hardness, the ionization parameter, the density and temperature of the ionized gas, and the heating mechanism of the gas \citep[e.g.,][]{Baldwin1981, Kaufman2006, Cormier2015, Cormier2019, Polles2019}. The IR fine-structure lines are less affected by extinction and dust attenuation \changes{than optical lines} and are therefore ideal for studying star-forming regions. 

Previous studies of mostly spatially unresolved infrared spectroscopy in low metallicity star-forming galaxies find harder radiation fields, extended, bright [\ion{O}{3}] emission, and an overall more porous structure \citep{Hunt2010, Cormier2015, Cormier2019}. \changes{The majority of spectra analyzed in these studies, however, are unresolved and contain emission from the different phases of the ISM and a variety of \hii\ regions that vary in properties (density, ionization parameter, stellar population, etc.) blended into one beam or observation.} Spatially resolved studies that focus on individual \hii\ regions, on the other hand, bring additional information and can be considerably more powerful at constraining physical conditions. Of course, they can only be carried out in nearby sources. The Small Magellanic Cloud (SMC), at one-fifth solar metallicity and only 63 kpc \changes{\citep{Dufour1984, Russell1992}} away provides the ideal laboratory for studying the low metallicity ionized gas in spatially-resolved detail.

This paper focuses on studying the ionized gas emission around the class of the hottest and most luminous stars in the local universe, a Wolf-Rayet (WR) star. After an O-type star loses its hydrogen-rich envelope (either from binary stripping and accretion, or via stellar winds removing the outer layers of the star), the inner, hot core is exposed and forms a WR star \citep{Crowther2006}. WR stars are characterized by effective temperatures of $T_{*}\sim$ 50 - 110 kK, luminosities of $\log L \sim 5 - 6.2 \lsun$, have strong stellar winds with velocities that range from $\rm v_{\infty} \sim 1500 - 5000$ \kms, and mass outflow rates of $\rm \dot{M} \sim 10^{-4} - 10^{-5} \msun \ yr^{-1} $ \citep{Crowther2006}. These stellar winds compress the surrounding material, forming a stellar wind blown bubble \citep[][]{Weaver1977}. Simultaneously, the bright ultraviolet (UV) flux from the WR star creates a highly ionized \hii\ region, forming a structure often called a \changes{WR nebula, a subclass of \hii\ regions}  \citep[e.g.,][]{Chu1981}. Since WR stars inject massive amounts of energy into the surrounding medium, they can have a profound influence on shaping the nearby ISM and energy exchange in \changes{stellar clusters} \citep[e.g.,][]{Sokal2016}.

\changes{This paper studies the WR nebula N76 that harbors one of the hottest, most luminous WR stars in the SMC \citep{Shenar2016}. First classified by \citet{Henize1956} as an \hii\ region, \citet{Garnett1991} identified broad He~II emission in N76, indicative of a WR nebula. N76 is powered by a well-studied binary system consisting of a WN4 and a O6 I(f) star \citep{Niemela2002, Shenar2016}. The goal of this work is to model the physical conditions of the ionized gas in N76 using mid and far infrared spectroscopy and the {\tt Cloudy} photoionization code \changes{\citep{Ferland2017}} to provide a detailed, resolved view of the impact the most energetic stars have on the metal-poor ISM. Many low metallicity dwarf galaxies that have been studied in the past \citep{Wu2006, Hunt2010, Cormier2015, Cormier2019} contain a large number of WR stars and are undergoing a burst of active star formation, but these galaxies are too far away to examine \newchanges{the specific impact of WR stars in detail}. Therefore N76 provides the ideal case study to isolate the impact WR stars have on the low metallicity ISM due to its proximity and simple, well characterized single ionization source.}

Observations of Milky Way WR nebulae show that they vary in morphology, but often appear as thin, bubble structures, or disrupted shells \citep{Chu1981, Chu1983, Toala2015}. In particular, the well-studied WR nebula NGC 6888 shows a double shell model with a denser inner shell ($\rm n_e \sim 400\,\cm$) and a thinner outer shell ($\rm n_e \sim 180\,\cm$) \citep{Fernandez2012, Rubio2020}. Due to the rarity of Wolf-Rayet nebulae \citep[e.g.,,][]{Conti1989, Hainich2014, Neugent2018}, there have been few observations investigating the properties of WR nebulae at low metallicities. It is possible that the lower dust abundance and harder radiation fields at low metallicities will allow high energy photons to penetrate deeper into the low metallicity ISM, forming larger, lower density nebulae with high ionizing photon escape fractions. 

Further, understanding how WR stars directly affect their surrounding environment and host galaxies is essential to deciphering the nature of extreme emission line galaxies (EELGs). These are local analogues to galaxies during the Epoch of Reionization, characterized by bright emission lines from high ionization species (such as [O III], C III], C IV, He II, and [\ion{Ne}{5}]), low metallicities, and highly ionized gas \citep[e.g.,][]{Atek2011, Maseda2013, Rigby2015, Senchyna2017M, Berg2019, Olivier2022}. The ionization source for the extreme emission found in these galaxies is unclear, but ultraluminous X-ray sources and radiation from WR stars are possible explanations. \changes{The observations of N76 presented in this work will directly test whether low metallicity WR nebula can produce the high ionization species ([\ion{O}{4}] and [\ion{Ne}{5}]) associated with EELGs.}

\changes{Below we outline the sections of this paper.} In Section \ref{sec:observations}, we describe the \spitzer\ and \herschel\ observations and the methods of producing line-integrated spatially resolved emission line maps. Section \ref{sec:models} describes the {\tt Cloudy} photoionization model input conditions and N76 geometry. We then compare the photoionization models to the spatially resolved IR data in Section \ref{sec:results}. We discuss the implications of these results, putting N76 in context with other low metallicity galaxies and WR nebulae in Section \ref{sec:discussion}. Lastly, our conclusions are described in Section \ref{sec:conclusions}.

% Further, understanding the nature of how WR stars affect their surrounding environment and host galaxies is essential in deciphering the nature o Wolf-Rayet galaxies behave. These galaxies  

% Detailed, resolved measurements of individual star-forming regions are needed to fully understand  

% We are particularly interested in studying the gas most adjacent to a given massive star: the ionized medium in an \hii\ region. 

\section{Observations}
\label{sec:observations}
% \subsection{Infrared Observations}
% \begin{itemize}
%     \item \underline{Figure:} N76 in context with the SMC, RGB image background with the various orders' coverage
%     \item Mapping strategy (reference \citet{Sandstrom2012}) and description of the different orders
% \end{itemize}

\subsection{Spitzer IRS observations and spectral line images}
\label{sec:linemaps}
% \begin{itemize}
%     \item Description of CUBISM \citep{cubism} to create the spectral cubes with a focus on diffuse line emission 
%     \item Used PAHFIT \citep{Smith2007} to fit the emission lines with a thermal dust continuum component. Fit the spectrum pixel by pixel and each individual order separately 
%     \item Subsection on how the PACS linemaps were created
%     \item \underline{Figure:} maps of all the emission lines found in N76.
%     \item \underline{Figure:} spectrum/fit example?
% \end{itemize}

The \spitzer\ data come from the InfRared Spectrograph (IRS) as part of the \spitzer\ Spectroscopic Survey of the Small Magellanic Cloud (S$^4$MC). The full observations are described by \citet{Sandstrom2012}, but we summarize them briefly here. We downloaded the raw Basic Calibrated Data (BCD) files from the \spitzer\ Heritage Archive from project GO 30491 (PI: A. Bolatto) with pipeline version S18.18.  We use the long-low (LL) and short-low (SL) modules of the IRS to cover the 5.2 \micron\ - 38.4 \micron\ range with a resolving power ranging from $R\sim 60-120$ \citep[coverage of the modules is shown in Figure 3 of][]{Sandstrom2012}. The spatially resolved maps are produced by stepping the IRS slit perpendicular and parallel to the source in steps of half a slit width for the LL module (5\arcsec.08) and the full slit width for the SL module (3\arcsec.7). The LL maps for N76 had $75 \times 6$ steps ($376\arcsec \times$ 395\arcsec ) with a 14~s integration time per position. The SL maps are created similarly, except the steps are using the full slit width in order to increase coverage and have $120 \times 5$ steps ($220\arcsec \times$ 208\arcsec ). 

Each set of \change{LL and SL} observations has an associated background observation taken at R.A. $\rm 1^h 9^m 40^s$ Dec. $-73^{\circ} 31\arcmin 30\arcsec$ (J2000), a region chosen from the Multiband Imaging Photometer for Spitzer (MIPS) and Infrared Array Camera (IRAC) observations of the SMC to have negligible emission from the SMC itself \citep{Bolatto2007, Leroy2007, Sandstrom2010}. We subtract the background from the BCD files, which removes emission from the zodiacal light of the Milky Way cirrus, and reduces the number of ``hot''  pixels that contaminate the IRS data. 

\subsubsection{Constructing the data cubes}

The BCD files are assembled into cubes using the software CUBISM \citep[we use version 1.8]{cubism}. CUBISM processes the two-dimensional slit spectral images created by scanning the IRS slit across the source into three-dimensional spectral data cubes through polygon-clipping-based re-projection. This algorithm is flux-conserving and is specially adapted for the IRS on \spitzer. \change{We apply a slit-loss correction for extended sources which assumes uniform extended emission inside and outside the slits and corrects the native tuning of the IRS pipeline that assumes point sources.} The IRS is susceptible to hot, ``rogue'', or bad pixels that \newchanges{appear} on the cube images as repeating stripes, resulting from the bad pixel scanning across the image through the grid of positions. These pixels are often caused by interactions between the IRS CCD and solar wind particles. The positions and intensity of the rogue bad pixels can vary on scales of hours to days, making them unpredictable. Some bad pixels can be flagged automatically through CUBISM, but many must be flagged by hand. We perform extensive flagging of the bad pixels in each data cube ($\sim~7\%$ of data), with particular focus on the emission from the spectral lines we plan to model, in order to produce the cleanest spectral line maps. 

\begin{figure*}
    \centering
    \includegraphics[width=\textwidth]{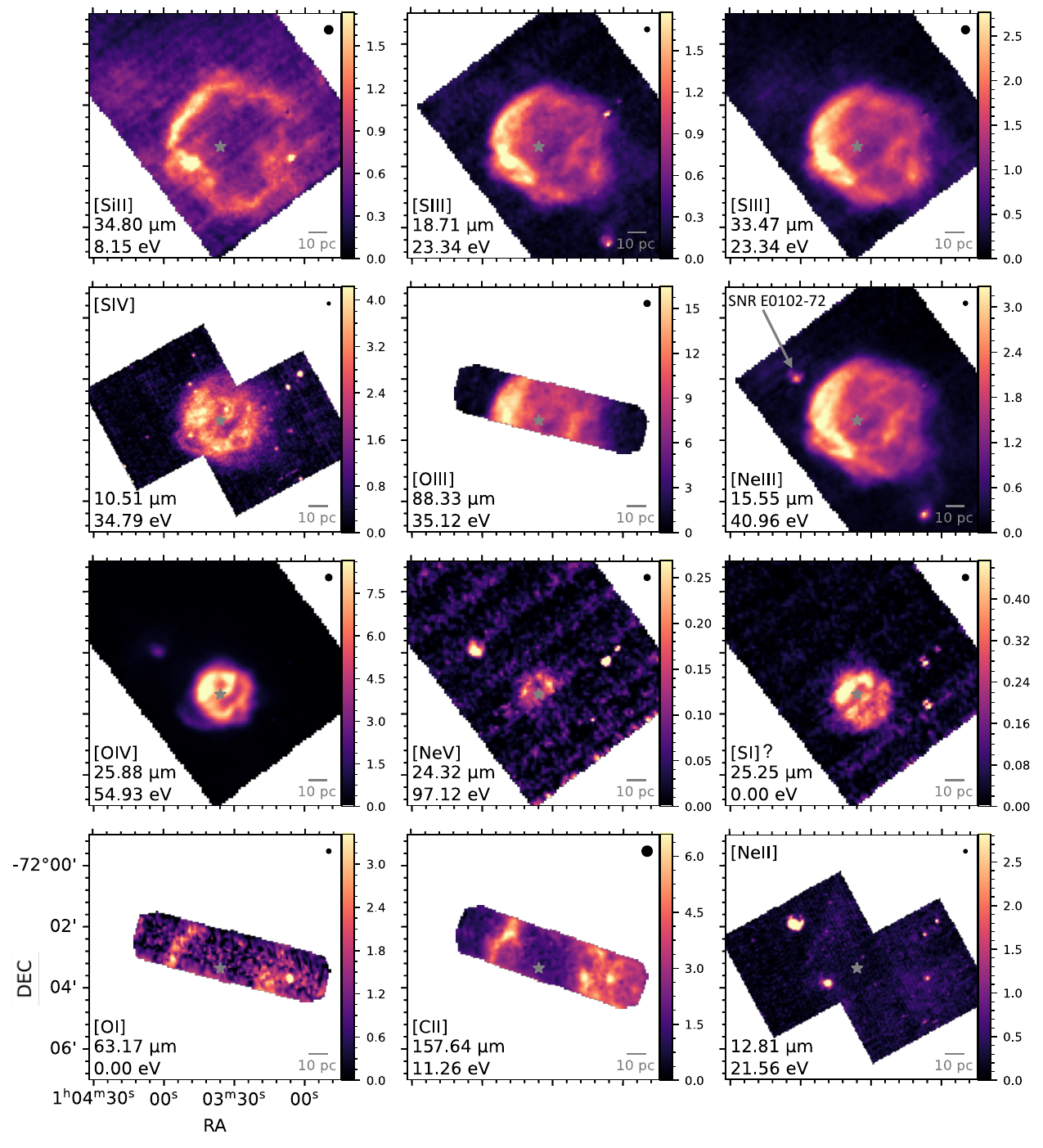}
    \caption{Images of infrared line emission in N76 where the color scale corresponds to the line brightness in $\rm 10^{-8} \ W \ m^{-2} \ sr^{-1}$. The location of the WR-O star binary AB7, the ionization source of N76, is illustrated by a gray star in the center of each image. The supernova remnant (SNR) E0102-72 is located NE, on the upper left of N76, and is seen in some of the emission lines ([\ion{O}{4}], [\ion{Ne}{5}], [\ion{Ne}{3}], [\ion{Ne}{2}]). The full width at half maximum (FWHM) of the point spread function (PSF) is shown in the upper right corner of each panel, and the ionization potential of the ion is displayed in the lower left corner together with the wavelength of each transition. These images show how the structure of N76 changes depending on the ion. }
    \label{fig:linemaps}
\end{figure*}

\subsubsection{Fitting spectral line maps}

We use the program PAHFIT \citep{Smith2007} to create line maps for each emission line in the IRS bandpass. PAHFIT uses a physically motivated model to simultaneously fit multiple components in an IRS spectrum. Its model includes dust continuum in fixed equilibrium temperature bins, starlight, bright emission lines, individual and blended PAH features, and extinction from silicate grains. In our fits we do not include extinction because the SMC has very low levels of mid-IR dust absorption, due to its \newchanges{relatively} low gas-to-dust ratio \citep{Lee2009, Roman-Duval2014}. The mapping strategy produces maps for both the SL and LL modules, but the placement of \change{the apertures corresponding to each spectral order} creates an offset of the mapping area between the different modules and orders, resulting in different coverage for SL1, SL2, LL1, and LL2 \citep[see Fig.\ 3 in][for an illustration of the coverage on N~76]{Sandstrom2012}. To maximize the area of the line maps, we fit each IRS module and spectral order separately with PAHFIT. 

% The foreground extinction from the Milky Way dominates \citep[][]{Gordon2003} and is subtracted from our data from the background observation. 

In order to create the emission line map, we first extract the spectrum for each pixel in the data cube. \changes{The native pixel size for the LL cubes is 5.07\arcsec , which is \newchanges{smaller than} the PSF at the lowest wavelength line ([\ion{S}{3}] 18 \micron\ , PSF FWHM of 4.15\arcsec ) and highest wavelength ([\ion{Si}{2}], PSF FWHM of 7.61\arcsec ). The SL native pixel scale is 1.85\arcsec , which samples the PSF of the lowest wavelength line ([\ion{S}{4}], PSF FWHM of 2.40\arcsec ) and the highest wavelength line ([\ion{Ne}{2}], PSF FWHM of 2.83\arcsec ).} We pass the spectrum into PAHFIT with a custom emission line fitting list to ensure the fainter lines not included in the default fitting list are incorporated. Then we extract the result of the line fit and construct the resulting line map using the fits for the individual pixels. The line maps of all emission lines (excluding PAHs and quadrupole H$_2$ transitions) found in N76 are presented in \autoref{fig:linemaps}. \changes{The integrated spectrum of N76 is presented in \autoref{fig:spec}.}

\changes{In addition to the N76 nebula, the IRS maps also include emission from a well-studied nearby core-collapse supernova remnant SNR E0102-72 \citep{Hayashi1994, Blair2000, Chevalier2005}. The focus of this work is on N76, so we refer the reader to \citet{Sandstrom2009} which investigates the mid-infrared properties of SNR E0102-72 in detail with the \spitzer\ IRS dataset also used in this work. The SNR is masked out for all analysis in this paper. }

\subsection{Herschel PACS observations and spectral line images}
\label{sec:PACS}
In addition to the \spitzer\ IRS data, we also report far-infrared observations from the \herschel\  Photodetector Array Camera and Spectrometer (PACS) project 1431 (PI: R. Indebetouw). We downloaded the data from the ESA \herschel\ Science Archive and use the level 2 pipeline reduced data, which contains spectral data cubes of the [\ion{C}{2}] 158 \micron, [\ion{O}{1}] 63 \micron, and [\ion{O}{3}] 88 \micron\ lines. To produce the spectral line maps, we fit a polynomial of order one to the baseline, remove the continuum, and integrate over the spectral line. This process was repeated pixel-by-pixel over the full datacube to produce a line intensity image for [\ion{C}{2}], [\ion{O}{1}], and [\ion{O}{3}]. Lastly, we convert to the same surface brightness units $\rm 10^{-8} \ W \ m^{-2} \ sr^{-1}$ used with the IRS. We compared this reduction to an independent reduction using the procedure in \citet{Cormier2015} and found excellent agreement within the uncertainties (priv comm, D. Cormier).

\subsection{Estimation of uncertainty in emission line surface brightness}
\label{sec:unc}

For the \spitzer\ IRS data, CUBISM provides a uncertainty cube that corresponds to the statistical uncertainty of the IRS measurement. These uncertainties are then propagated into the PAHFIT code, where they are used for the chi-squared minimization fitting. The resulting uncertainty from the PAHFIT output is propagated into our emission line images. These estimates of the uncertainties, however, are lower limits since they only include noise from the detector readout and do not comprise any systematic errors \changes{such as uncertainties in the absolute calibration or the effective PSF due to perpendicular slit offsets in the map}. 

For the \herschel\ PACS data, we compute the RMS noise outside of the signal integration per each pixel as a 1$\sigma$ statistical uncertainty for each channel. We then propagate the channel \changes{RMS} to calculate an uncertainty for the integrated signal. This process is repeated per pixel to create a map of the statistical uncertainties for [\ion{C}{2}], [\ion{O}{1}], and [\ion{O}{3}]. 

In addition to the statistical uncertainties, we also calculate a ``background'' level for each emission line. We concentrate on modeling the emission from the \hii\ region created by the Wolf-Rayet/O star binary ionization source AB7. There are other ionization sources across the SMC that can produce diffuse ionized gas throughout the galaxy \citep[e.g.,][]{Haffner2009}. In order to quantify and remove this extraneous emission, we identify a blank part of the map that is not contaminated by emission from the \hii\ region (typically $\sim$80~pc away from AB7). We calculate the median surface brightness in this patch of sky and consider it the ``background'' for the given emission line. The background is very low, typically $\lesssim 5\%$ of the emission line intensity. \changes{The background levels are generated for the IRS maps only because the PACS field of view only covers the N76 nebula and there are no regions in the PACS maps that would not be contaminated by N76.}

% These estimates of the uncertainties, however, are only statistical and do not include systematic errors. In order to estimate the systematic errors, in each line map we find a blank part of the sky that is not contaminated by emission from the \hii\ region. We calculate the median surface brightness in this patch of sky and consider it the ``background'' for the given emission line. This background is subtracted on all radial profile plots in order to compensate for the unknown systematics. 

% \begin{figure*}
%     \centering
%     \includegraphics[width=\textwidth]{N76_fullspec_reg_integrated.pdf}
%     \caption{Example spectrum showing the ionized gas lines in N76 we model with {\tt Cloudy}, at one line of sight towards 01h03m40s -72d03m20s in a singular PSF-matched pixel aperture. Line labels are in blue and the orders of the IRS instrument are presented through the background color. This position corresponds to the peak of the faintest lines in the N76 nebula, [\ion{Ne}{5}] and [\ion{S}{1}], and is close to AB7 in an area tracing the high ionization lines.}
%     \label{fig:spec}
% \end{figure*}

\begin{figure*}
    \centering
    \includegraphics[width=\textwidth]{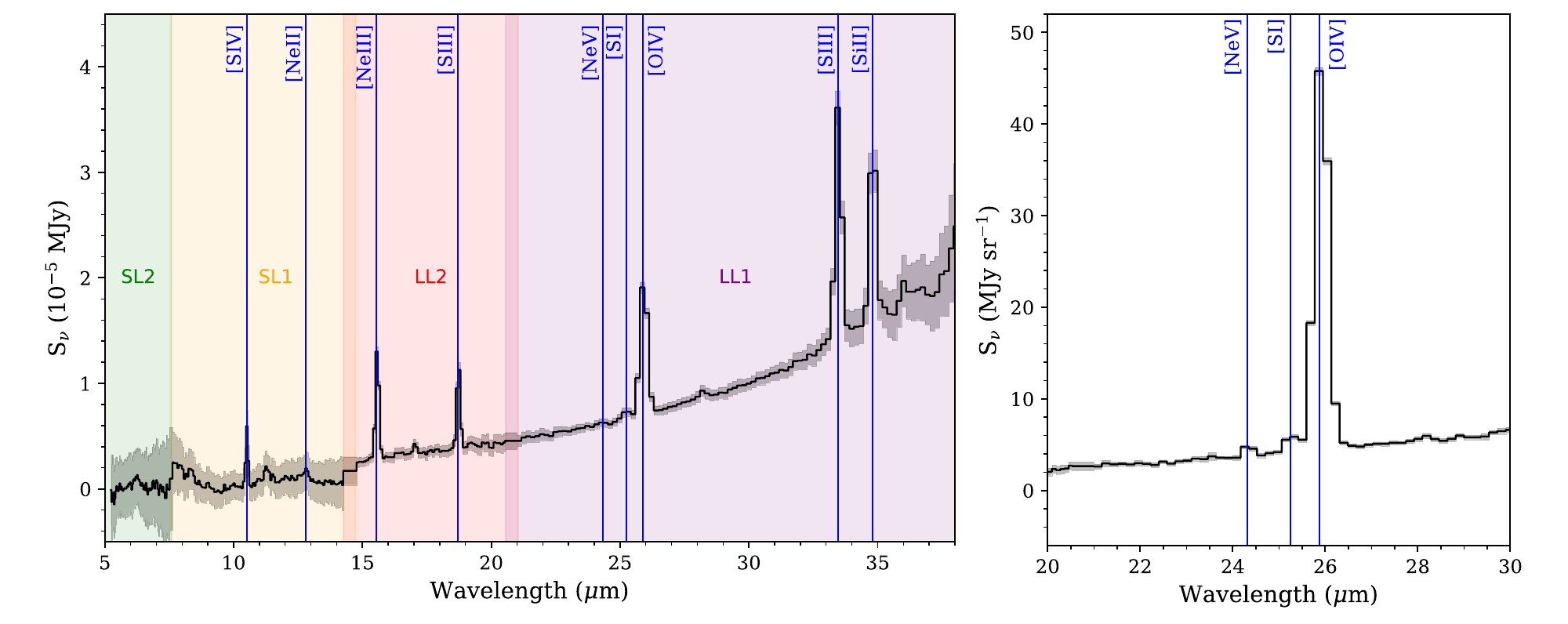}
    \caption{\changes{Left: Integrated spectrum of the N76 region showing the ionized gas lines in N76 we model with {\tt Cloudy}. Line labels are in blue and the orders of the IRS instrument are presented through the background color. Right: spectrum at a single line of sight towards the location of the WR binary AB7 that highlights the [\ion{Ne}{5}] and possible [\ion{S}{1}] features.}}
    \label{fig:spec}
\end{figure*}

\subsection{PSF matching}
\label{sec:psf_match}

The spatial resolution of the \spitzer\ IRS changes as a function of wavelength as the telescope is diffraction limited, with a point spread function (PSF) full width at half maximum (FWHM) ranging from 7.6\arcsec at 34.8~\micron\ for [\ion{Si}{2}] to 2.4\arcsec at 10.5~\micron\ for [\ion{S}{4}]. In order to properly model the spectra, we need to convolve the spectral line images to a common PSF \citep[e.g.,][]{Smith2009, Sandstrom2009}. We follow the procedure described by \citet{aniano2011} to create custom convolution kernels that will transform an image with a narrower PSF to a broader PSF. We refer the reader to \autoref{sec:app-psf} for a detailed description of the PSF matching. 

\subsection{Radial profiles}
\label{sec:rad_prof}
The spatial information given by the resolved infrared emission lines helps differentiate between photoionization models. \changes{{\tt Cloudy} simulates \hii\ regions with a 1-dimensional (1D) radiative transfer code, but can accommodate a variety of \hii\ region geometries with the assumptions in a given model. Due to the spherical nature of N76 (see \autoref{fig:linemaps}), we use the spherical geometry in {\tt Cloudy} to best match the observations.} In order to properly compare with the models, we average the spectral line images to create surface brightness radial profiles. N76 is an ideal region for this method because it is approximately spherically symmetric and mostly isolated from other bright regions (see \autoref{fig:linemaps}). We use the location of AB7 as the center of N76 and compute an azimuthal average with a radial bin size corresponding to one pixel or $5\arcsec$. 
%In order words, these profiles are defined as the radially averaged surface brightness as a function of distance from AB7. 
The radial profiles show how the line emission varies with distance away from the ionization source, thereby preserving some of the spatial information from the spectral line images but making it simple to compare to properly integrated 1D {\tt Cloudy} models. The radial profiles measured for each line are presented in \autoref{fig:radprof_den}, showing that higher ionization lines generally peak closer to AB7. We also report the standard deviation in a given azimuthal bin as the gray uncertainty bars in the profile.

\subsection{Integrated Intensities}
In addition to examining the resolved surface brightness of the infrared emission lines through radial profiles ($\S$\ref{sec:rad_prof}), we also record the total intensities by integrating across the nebula for the emission lines from \spitzer\ IRS. \changes{We aim to compare the observed total intensity of N76 for each line to the predictions of the total intensity from {\tt Cloudy} as a secondary check of the models. The lines from \herschel\ PACS ([\ion{C}{2}], [\ion{O}{1}], and [\ion{O}{3}]), however, only cover a strip of the N76 nebula and not correcting for this missing area and flux leads to under-predictions of the total intensity of N76. We identify the missing area for each PACS line by using the \spitzer\ IRS emission line that has the most similar radial profile. For [\ion{C}{2}] and [\ion{O}{1}] we use [\ion{Si}{2}], and for [\ion{O}{3}] we use [\ion{Ne}{3}]. To correct for the missing flux, we multiply the flux contained in the PACS footprint by the ratio of total area of N76 for that emission line to the area of the PACS footprint. We note that this modified version of the PACS maps are only used for the integrated total intensities reported in \autoref{tab:lines} and are not used for the radial profiles.} The uncertainties in the integrated intensities are derived from the statistical uncertainties propagated through the total intensity calculation (see $\S$\ref{sec:unc}). The integrated intensities for all the emission lines are reported in \autoref{tab:lines}. 

\changes{\subsection{[\ion{O}{4}] and [\ion{Fe}{2}] Line Blend \label{sec:blend}} The spectral resolution of R$\sim$70 in the \spitzer\ IRS LL module causes the [\ion{O}{4}] 25.89 \micron\ and [FeII] 25.99 \micron\ lines to be blended. For this work, we assume that the flux of the line observed at 25.9 \micron\ is dominated by [\ion{O}{4}] emission. \newchanges{The {\tt Cloudy} models are tuned to the parameters of the AB7 system determined from modeling the optical and UV spectra in \citet{Shenar2016} (see \S \ref{sec:models}) and find that [\ion{O}{4}] 25.89 \micron\ emission is two orders of magnitude brighter than [\ion{Fe}{2}] 25.99\micron\ line, strongly indicating that [\ion{Fe}{2}] makes a negligible contribution to the 25.9 \micron . The {\tt Cloudy} models are inconsistent with a large contribution of [\ion{Fe}{2}] to the 25.9 \micron\ blend due to the nature of hot WR star as the ionization source.} Further, the resolved images of N76 show that the emission at 25.9 \micron\ is centrally concentrated and associated with the high ionization zone of the \hii\ region, making it much more likely to be a line with an ionization potential of 55 eV ([\ion{O}{4}]) rather than 7.9 eV ([\ion{Fe}{2}]). \newchanges{There are also no detections of the iron lines at [\ion{Fe}{2}] 17.94 \micron\ or [\ion{Fe}{3}] 22.93 \micron\ in N76. \citet{Baldovin-Saavedra2011} use the high resolution \spitzer\ module (LH) which can resolve the 25.9 \micron\ [\ion{O}{4}] and [\ion{Fe}{2}] blend. In their observations of the Taurus Molecular Cloud, they find that the [\ion{Fe}{2}] 25.99 \micron\ line is a factor of three brighter than [\ion{Fe}{2}] 17.94 \micron , consistent with theoretical predictions in \citet{Hollenbach1989}. Using this relationship, we calculate an upper limit of the [\ion{Fe}{2}] 25.99 \micron\ in N76 of $1.3 \times 10^{-10} \ \rm W \ m^{-2} \ sr^{-1} $, which would imply $<5\%$ of the 25.9 \micron\ line surface brightness comes from [\ion{Fe}{2}] 25.99 \micron\ emission. Therefore, [\ion{Fe}{2}] contributes a negligible component of the 25.9 \micron\ line and we will assume for this work that the line flux is dominated by the [\ion{O}{4}] line. The [\ion{O}{4}] line dominating the 25.9 \micron\ blend is also consistent with observations of blue compact dwarf galaxies, which have similar metallicities and ionization parameters as N76 \citep{Hunt2010}.}

% Similar IRS spectra of Cassiopeia A that have a substantial [\ion{O}{4}] and [FeII] blend also present detections of the iron lines at [FeII] 17.94 \micron\ or [FeIII] 22.93 \micron\ \citep{Smith2009}. In N76, these lines are undetected, making it very unlikely that [FeII] contributes to the line at 25.9 \micron . Further, the resolved images of N76 show that the emission at 25.9 \micron\ is centrally concentrated and associated with the high ionization zone of the \hii\ region, making it much more likely to be a line with an ionization potential of 55 eV ([\ion{O}{4}]) rather than 7.9 eV ([FeII]). If the [FeII] line does have a significant contribution to the total flux at 25.9 \micron , then Cloudy models would substantially under-predict the flux associated with [\ion{O}{4}], which is not the case in N76 (see \S \autoref{sec:results}). Therefore, we are confident in declaring that the line at 25.9 \micron\ is predominately from [\ion{O}{4}] emission.}

\startlongtable 
\startlongtable 
\begin{deluxetable*}{cccccccccc} 
\tabletypesize 
\footnotesize 
\tablecaption{Line properties and intensities} 
\tablehead{ & & \multicolumn{1}{c}{Ionization} & & & & \multicolumn{2}{c}{$\mathrm{ I_{obs} \ (10^{-15} \ W \ m^{-2})}$} & \multicolumn{2}{c}{$\mathrm{ I_{cloudy} \ (10^{-15} \ W \ m^{-2})}$} \\ 
\colhead{Line} &  \colhead{$\mathrm{ \lambda \ (\mu m)}$} & \colhead{Potential (eV)} & \colhead{$\mathrm{ n_{crit} \ (cm^{-3})}$} &  \colhead{FWHM ($\arcsec$)} & \colhead{Instrument} & \colhead{Total} & \colhead{$\mathrm{Orig}$} & \colhead{$\mathrm{n_{H} = 4 \ cm^{-3}}$} & \colhead{$\mathrm{n_{H} = 10 \ cm^{-3}}$}}
\startdata 
$\rm \left[SiII\right]$ & 34.80 &  8.2 & 2 $\times \ 10^{3}$ & 7.61 & LL1 & 14.75 $\pm$ 0.01 & $\ldots{}$ & 17.16 & 14.37 \\
$\rm \left[SIII\right]$ & 18.71 & 23.3 & 2 $\times \ 10^{4}$ & 4.15 & LL2 & 11.28 $\pm$ 0.73 & $\ldots{}$ & 13.00 & 12.49 \\
$\rm \left[SIII\right]$ & 33.47 & 23.3 & 7 $\times \ 10^{3}$ & 7.17 & LL1 & 18.93 $\pm$ 0.00 & $\ldots{}$ & 20.57 & 19.52 \\
$\rm \left[SIV\right]$ & 10.51 & 34.8 & 5 $\times \ 10^{4}$ & 2.40 & SL1 & 17.58 $\pm$ 2.35 & $\ldots{}$ & 35.05 & 41.65 \\
$\rm \left[OIII\right]$ & 88.33 & 35.1 & 5 $\times \ 10^{2}$ & 5.35 & PACS & 111.68 $\pm$ 2.25 & 50.21 $\pm$ 1.01 & 138.40 & 143.00 \\
$\rm \left[NeIII\right]$ & 15.55 & 41.0 & 3 $\times \ 10^{5}$ & 3.51 & LL2 & 22.05 $\pm$ 0.70 & $\ldots{}$ & 28.33 & 28.67 \\
$\rm \left[SI\right]$ & 25.25 &  0.0 & 1 $\times \ 10^{5}$ & 5.44 & LL1 & 1.04 $\pm$ 0.14 & $\ldots{}$ & 0.00 & 0.00 \\
$\rm \left[OI\right]$ & 63.17 &  0.0 & 9 $\times \ 10^{5}$ & 3.63 & PACS & 15.74 $\pm$ 3.37 & 4.49 $\pm$ 0.90 & 1.30 & 1.02 \\
$\rm \left[CII\right]$ & 157.64 & 11.3 & 5 $\times \ 10^{1}$ & 9.43 & PACS & 60.35 $\pm$ 1.20 & 16.70 $\pm$ 0.33 & 1.95 & 1.31 \\
$\rm \left[NeII\right]$ & 12.81 & 21.6 & 7 $\times \ 10^{5}$ & 2.83 & SL1 & 4.52 $\pm$ 1.70 & $\ldots{}$ & 0.92 & 0.71 \\
$\rm \left[OIV\right]$ & 25.88 & 54.9 & 1 $\times \ 10^{4}$ & 5.66 & LL1 & 20.35 $\pm$ 0.04 & $\ldots{}$ & 28.99 & 29.38 \\
$\rm \left[NeV\right]$ & 24.32 & 97.1 & 1 $\times \ 10^{5}$ & 5.22 & LL1 & 0.20 $\pm$ 0.06 & $\ldots{}$ & 0.49 & 0.37 \\
\enddata 
\tablecomments{Description of the spectral lines modeled, including the integrated intensities measured and the predicted integrated intensities with two densities from Cloudy. The Ionization Potential is the photon energy required to produce the observed ionization state from the lower ionization state. The critical density is for collisions with electrons. The lines observed with \textit{Herschel}/PACS, and given as $I_{obs}$ Orig, do not cover the full nebula (see \autoref{fig:linemaps}); $I_{obs}$ Total contains the total intensity corrected for this missing flux. \label{tab:lines}}
\end{deluxetable*}

\begin{figure*}
    \centering
    \includegraphics[width=\textwidth]{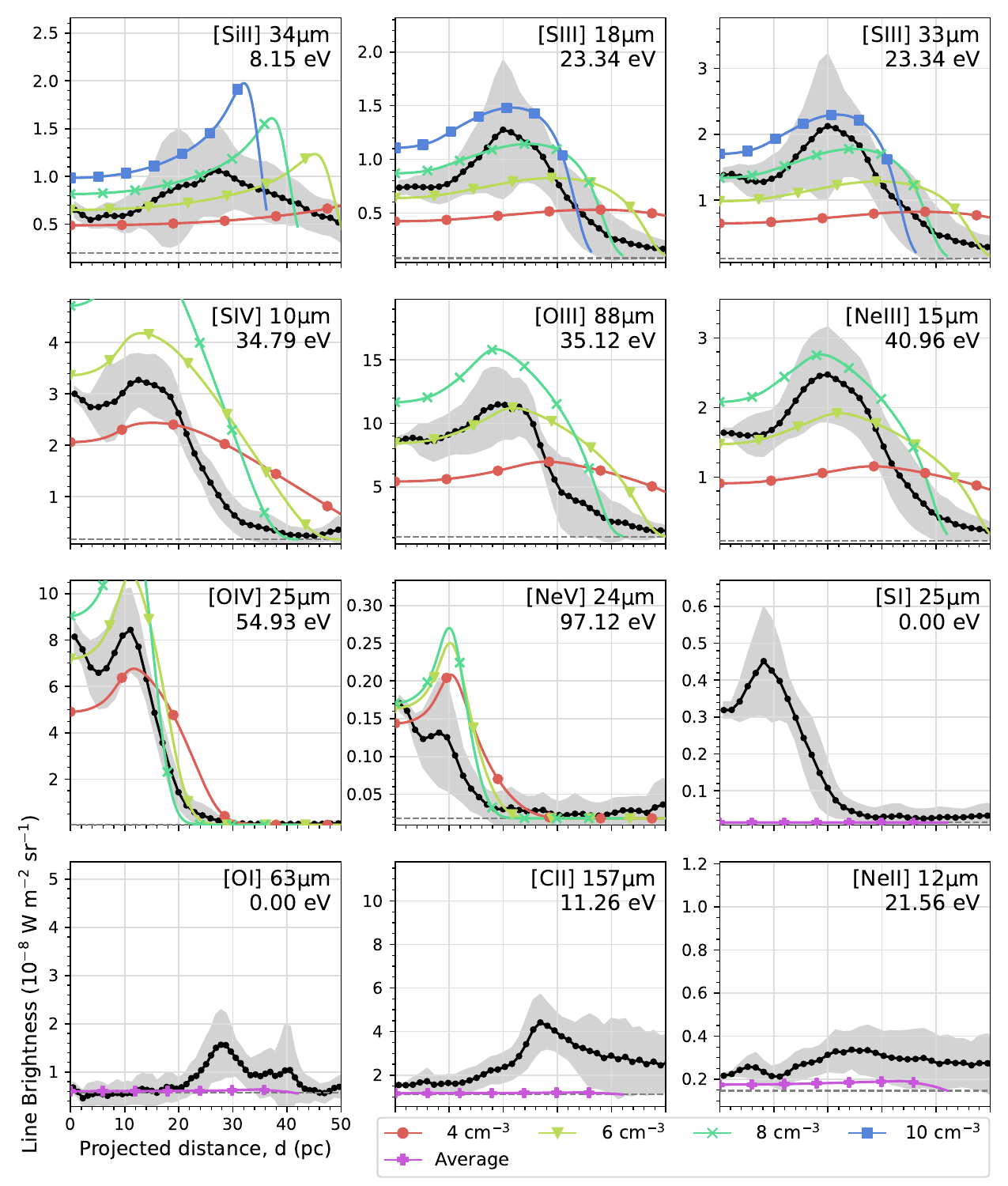}
    \caption{Radial profiles of observed emission line brightness and {\tt Cloudy} photoionization models. The black circles represent the average line brightness of a given emission line as a function of distance away from AB7, the gray region is the standard deviation in a given bin, and the dashed line represents the background diffuse contribution to the emission line (see $\S$\ref{sec:unc}). The {\tt Cloudy} models vary with hydrogen density and are represented by the colored lines and symbols, with the average of all models presented in purple crosses. The ions that originate from the photoionized gas (excluding emission lines that can be produced through neutral material ([\ion{C}{2}] and [\ion{O}{1}]) see section \ref{sec:result_neut}, or the diffuse ionized gas, $\S$\ref{sec:NeII}) are well predicted by the {\tt Cloudy} models that range in density from $\rm n_{H} \sim 4 \ cm^{-3} - 10 \ cm^{-3}$. Both the surface brightness and the shape of the radial profiles are well predicted in these cases.}
    \label{fig:radprof_den}
\end{figure*}

\section{Photoionization models}
\label{sec:models}

% Define the photoionization model

\subsection{{\tt Cloudy} model parameters}
\label{sec:parameters}
% \begin{itemize}
%     \item Set up and description of the {\tt Cloudy} model \citep{Ferland2017}. Describe the stopping conditions and the dust properties used
%     \item SED from PoWR (appropriate citation and parameters here)
%     \item Luminosity from \citet{Shenar2016}
%     \item Table of metallicities used (combination of a few different papers to account for depletion)
% \end{itemize}

We use version C17.01 of {\tt Cloudy} \citep{Ferland2017} to model the properties of the ionized gas in N76. {\tt Cloudy} is a spectral synthesis code designed to simulate physical conditions and resulting spectrum from gas and dust that is exposed to an ionizing radiation field. In order to produce a model, a variety of input parameters are required and listed below. 

\textit{Ionizing source SED:} The shape of the source spectrum as a function of photon energy must be specified in {\tt Cloudy}, which for N76 corresponds to the spectral energy distribution (SED) of the central binary. We use SED models in \citet{Shenar2016} that constrain the stellar parameters of AB7 binary through a combination of UV spectroscopy from the International Ultraviolet Explorer (IUE) and the Far Ultraviolet Spectroscopic Explorer (FUSE) as well as optical spectroscopy taken from \citet{Foellmi2003}. The stellar parameters of the WR and its O star companion were identified by matching the observed spectra to models from the non-LTE Potsdam Wolf-Rayet (PoWR) model atmosphere code\footnote{\url{http:// www.astro.physik.uni-potsdam.de/PoWR.html}} \citep{Grafener2002A&A...387..244G, Hamann2003, Sander2015, Hamann2004, Hainich2019}. \changes{These PoWR models are set to use the SMC metallicity and derive a variety of parameters for the WR and O star.} \citet{Shenar2016} report that the WR is a WN4 spectral type with a temperature of $T_* =  105000 \rm K$, wind speed of $v_{\infty} = 1,700$ \kms , surface gravity of log $g_* = 4.7 \  \rm cm \ s^{-2}$, a low ($X_{\rm H} \approx 0.15$) hydrogen fraction, a transformed radius of log\,$R_t = 0.75 \rm \ R_{\odot}$, and a mass loss rate of log $\dot{M} = -5.0 \rm \ M_\odot \ yr^{-1}$ . The O star companion is classified as an O6 I(f) star and has a temperature of $T_* =  36$ kK, wind speed of $v_{\infty} = 1,500$ \kms , surface gravity of log\,$g = 3.6$ $\rm cm \ s^{-2} $, and a mass loss rate of log $\dot{M} = -7.0 \rm \ M_\odot \ yr^{-1}$. 

\changes{The PoWR models are normalized \newchanges{by default} to a luminosity of $\rm L_{\odot} = 5.3 \lsun$.} Therefore, we scale the SEDs fitted in \citet{Shenar2016} by their luminosities of $\rm log \ L_{\odot}  = 6.1$ for the WN4 and $\rm log \  L_{\odot} = 5.5$ for the O6 components. We then add the luminosity weighted WN and O6 SEDs together to produce a total SED for the AB7 binary system, shown in \autoref{fig:sed}. The feature at 40~eV in the spectrum is due to a strong FeV emission line at 305.31\,\AA.

\begin{figure}
    \centering
    \includegraphics[width=\columnwidth]{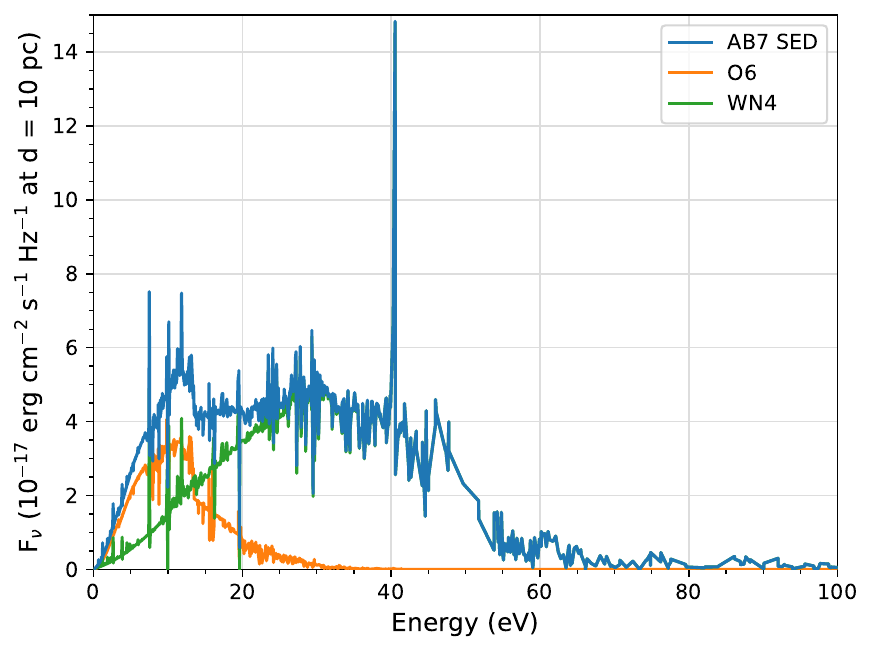}
    \caption{The SED of the AB7 system for {\tt Cloudy} input. We use the Potsdam Wolf-Rayet database models for the WN Wolf-Rayet \citep{Hamann2004, Todt2015} and O6 \citep{Hainich2019} stars in the AB7 binary and combine them into one input SED. The full SED for AB7 is in blue, the WR is green, and the O star is orange.}
    \label{fig:sed}
\end{figure}

%\textit{Luminosity of ionizing source:} The SED sets the shape of the ionizing source, but not the luminosity. We specify the luminosity of AB7 using values derived by \citet{Shenar2016}, with a value of $\rm log(L_{\odot}) = 6.1$ for the Wolf-Rayet and $\rm log(L_{\odot}) = 5.5$ for the O6 star.  

\textit{Elemental abundances:} Instead of scaling gas-phase Milky Way abundances to the metallicity of SMC, we employ the best gas-phase abundances available for SMC \hii\ regions. Relative abundances of elements are different in the Magellanic Clouds and the Milky Way due to different nucleosynthesis histories \citep{Russell1992}. For the majority of elements, we adopt the abundances given for \hii\ regions provided in \citet{Dufour1984} and \citet{Russell1992}, but we slightly modify the abundances for sulphur and silicon to best match the observations in N76 \changes{(see below)}. Our adopted abundances and the references used are in \autoref{tab:abun}. 

We decrease the sulphur abundance by a factor of two from the $\rm (S/H) = 3.9 \times 10^{-6}$ value reported in \citet{Russell1992}. Without this modification, the three sulphur lines observed in this work ([\ion{S}{3}] 18 \micron , [\ion{S}{3}] 33 \micron\ and [\ion{S}{4}] at 10 \micron ) are consistently over-predicted by a factor of two in our Cloudy models. This small variation in the sulphur abundance is also consistent with the scatter presented in \citet{Russell1990},  the value of $\rm (S/H) = 1.8 \times 10^{-6}$ reported in \citet{Dennefeld1983}, \changes{and the sulphur abundances found in \citet{Vermeij2002} and \citet{Lebouteiller2008} that use infrared data. }

Identifying an abundance for silicon is more complex because \newchanges{silicon} is often depleted in dust. While the dust abundance is often lower in \hii\ regions \newchanges{than HI regions or molecular clouds}, dust can still be present and have an impact on \hii\ regions, especially towards the outer edges where the ionized hydrogen fraction is lower and the [\ion{Si}{2}] emission is brightest. \citet{Russell1992} provide only a photospheric abundance for silicon that does not take into account the depletion from dust. We therefore use elemental depletion studies in the SMC to calculate the median gas phase abundance for silicon \citep{Tchernyshyov2015, Jenkins2017}. The median silicon depletion from these studies is 0.4 which \changes{happens to} accurately match the observed flux of [\ion{Si}{2}]. We note that this systematic uncertainty in the silicon \hii\ region abundance leads to an uncertainty in the prediction for the [\ion{Si}{2}] flux (see discussion in $\S$\ref{sec:result_neut}). 

\startlongtable 
\startlongtable 
\begin{deluxetable}{ccc} 
\tabletypesize 
\footnotesize 
\tablecaption{Adopted SMC Abundances  \label{tab:abun} } 
\tablehead{ 
\colhead{Element} &  \colhead{Abundance (X/H)} & \colhead{Reference}}
\startdata 
He & 8.13 $\times \ 10^{-2}$ & 2 \\
C & 1.45 $\times \ 10^{-5}$ & 1 \\
N & 4.27 $\times \ 10^{-6}$ & 2 \\
O & 1.07 $\times \ 10^{-4}$ & 2 \\
Ne & 1.86 $\times \ 10^{-5}$ & 2 \\
Al & 2.51 $\times \ 10^{-6}$ & 2 \\
Si & 4.88 $\times \ 10^{-6}$ & 3,4 \\
S & 1.95 $\times \ 10^{-6}$ & 2 \\
\enddata 
\tablecomments{References are (1) \citet{Dufour1984} (2) \cite{Russell1992}, (3) \cite{Tchernyshyov2015}, and (4) \citet{Jenkins2017}.}
\end{deluxetable}

\textit{Stopping criteria:} We are interested in modeling the ionized gas properties of N76 and \changes{therefore will run the {\tt Cloudy model} to the end of the ionization zone at the transition between the \hii\ region and PDR. We therefore define the HII region end to be the point at which the PDR begins and the gas is neutral. The focus of this work is to model the ionized gas so continuing the model into the PDR is unnecessary, but has the consequences of underestimating the emission from the neutral dominated lines [\ion{Si}{2}], [\ion{C}{2}], and [\ion{O}{1}] (see $\S$\ref{sec:result_neut}). After using a range of stopping conditions we find that an electron temperature cutoff of 2000~K accurately probes the ionized gas to the end of the ionization front.} 

\textit{Geometry}: We assume a spherical geometry to approximate the geometry of N76. The sphere command in {\tt Cloudy} sets the geometry to be closed and assumes that the gas fully covers the ionization source. 

\textit{Inner radius}: The inner radius is defined as the distance between the gas and the ionization source. \changes{The massive, fast  stellar winds produced from Wolf-Rayet stars often form a stellar wind-blown bubble (SWB) structure in the resulting nebula \citep[e.g.,][]{Weaver1977}. As we discuss in \S\ref{sec:wind}, the region closest to AB7 has been shocked by the stellar winds and is occupied by this tenuous very hot plasma.} For the purposes of our modeling, this effect is best reproduced by a central cavity in the gas. We vary the gas inner radius and find a radius of 10~pc to best reproduce the observed emission lines radial profiles. 

\textit{Hydrogen density}: The density input in {\tt Cloudy} is the total hydrogen density. \changes{We assume the density of the \hii\ region is constant, a reasonable assumption for classical, spherical \hii\ regions, and the density remains constant up to the ionization front of the region. We produce multiple constant densities models that vary from $\rm n_H = 1 - 50 \ \cm$ in order to best constrain the density in N76 with the observed emission lines.} The range of hydrogen densities between $\rm n_H = 4 - 10 \ \cm$ that fit the data best are presented in this work. 

Since the binary at the center of N76 is well-studied, most of the inputs in the {\tt Cloudy} model are using {\em a priori} knowledge of the ionization source, abundances, and geometry. 

% This demonstrates the power of photoionization modeling to accurately predict the resulting emission lines from a nebula. 

\subsection{Calculating predicted radial profiles}
\label{sec:proj_prof}
% \begin{itemize}
%     \item Description of how to go from the opacity of a line to the predicted surface brightness 
%     \item detailed description of the geometry (below)
% \end{itemize}

\begin{figure}
    \centering
    \includegraphics[width=\columnwidth]{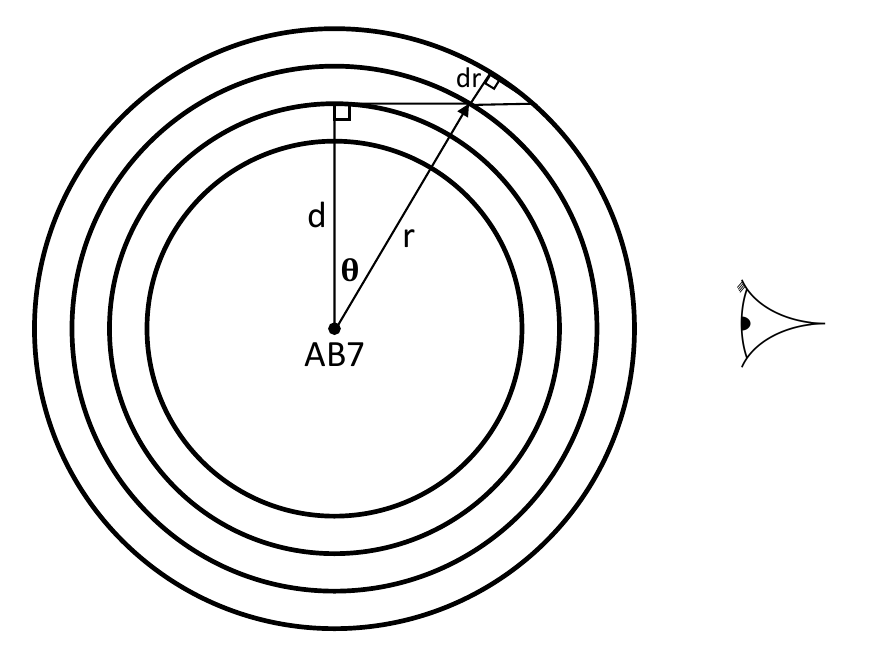}
    \caption{Schematic showing the geometry of calculating the projected intensity from {\tt Cloudy} models where $d$ is the projected distance, $r$ is the physical distance, and $dr$ is the differential distance between shells. The observer is peering from the right of the figure.}
    \label{fig:proj}
\end{figure}

N76 is an approximately spherical shell, also known as a bubble nebula, with a radius of about 40~pc. Because of the approximately symmetric nature of the nebula, we compare the {\tt Cloudy} models and the observations using radial profiles. {\tt Cloudy} models report the volume emissivity of each transition as a function of depth into the nebula. In other words, they provide the radial profile of emissivity in a sphere. To compare with the observations, we need to appropriately project these profiles to compute the azimuthal average surface brightness as a function of projected distance from AB7 for each of the emission lines. A point that is at a distance $d$ from the source in projection has contributions from shells that have radii between $d$ and infinity in proportion to their emissivity and projected thickness. \autoref{fig:proj} shows the geometry of the N76 {\tt Cloudy} model and the relationship between different shells, the projected distance $d$, and the physical distance $r$. The projected surface brightness for the {\tt Cloudy} model then is:

\begin{equation}
\label{eq:proj1}
    I(d) = \frac{1}{2 \pi} \int_{d}^{\infty} \frac{\epsilon(r)}{\sin \theta} dr \ , 
\end{equation}

\noindent where $r$ is the physical 3D distance from AB7, $\epsilon(r)$ is the radial dependence of the line volume emissivity reported from {\tt Cloudy}, $d$ is projected distance from AB7, and $\theta$ is the angle between $d$ and $r$. We convert \autoref{eq:proj1} to eliminate $\theta$ resulting in 

\begin{equation}
\label{eq:proj2}
    I(d) = \frac{1}{2 \pi} \int_{d}^{\infty} \frac{\epsilon(r)}{\sqrt{1 - d^2/r^2}} dr\ , 
\end{equation}

\noindent and calculate the surface brightness predicted from the {\tt Cloudy} model, $I(d)$, as a function of the projected distance and volume emissivity reported from {\tt Cloudy} in order to compare to the observed data. \changes{We set the {\tt filling factor} in {\tt Cloudy} to unity to reflect that the \newchanges{resolved emission in N76 is not clumpy.}}

The {\tt Cloudy} models use AB7 as the only ionization source, but there may be contributions of low ionization ionized gas from the warm ionized medium to our observed emission lines \citep[e.g.,][]{Haffner2009}. \newchanges{To account for this, we add the constant surface brightness background value calculated in $\S$\ref{sec:unc} to the radial profiles derived from the Cloudy models.} This background is typically very low representing $\lesssim 5\%$ of the total surface brightness. Lastly, the final {\tt Cloudy} profiles are convolved with a Gaussian kernel with a FWHM of 12\arcsec\ to make an equal comparison to the 12\arcsec\ PSF matched emission line images.  

\section{Results}
\label{sec:results}

\subsection{Comparison of models to observations}
\label{sec:results_model}
% \begin{itemize}
%     \item Add table of all the spectral lines used/fitted and there total intensities (maybe put in section 2 instead?)
%     \item \underline{Figure:} radial profiles with data and model comparison, in a similar arrangement to the spectral line maps
%     \item Line by line comparison of how well the profiles fit
%     \item Main conclusion: a simple, single density model that ranges from $8 - 16$ \density\ is consistent with the majority of observations 
%     \item The intensity alone of lines does not determine the ideal model as well as the shape of the profile (possible \underline{figure}?)
%     \item Qualitative description of the structure in the images - the tail/swirl seen in the [\ion{O}{4}] and [\ion{S}{4}] data, the brighter right side and diffuse left, diffuse emission throughout 
% \end{itemize}

\autoref{fig:radprof_den} shows the results of the constant density {\tt Cloudy} model with hydrogen densities $\rm n_{H} = 4 - 10$~cm$^{-3}$. We compare the models projected on the plane of the sky (\S\ref{sec:proj_prof}) with the measured emission line radial profiles (\S\ref{sec:rad_prof}). Therefore we compare not only the integrated intensity of the emission line, but also how well its predicted distribution matches the data.  The emission lines we sample are at a variety of ionization energies and therefore trace the different ionization zones in the nebula. For example, the [\ion{O}{4}] line ($\rm E_{ion} = 55 \ eV$) peaks at 10 pc while the [\ion{S}{3}] line ($\rm E_{ion} = 23 \ eV$) peaks at 20 pc. A model that predicts an emission line well will have a similar peak, spatial gradient, and ending position as the observed radial profile

There is not a single constant density model that is able to predict the shape and intensity of all of the emission lines, but most emission lines are consistent \changes{with any given} constant density model in the range of $\rm n_{H} = 4$~cm$^{-3}$ to $\rm n_{H} =10$~cm$^{-3}$. In some cases, the spatial distribution of the profile is better matched by a model that does not predict the intensity as well. For example, in the [\ion{S}{4}] profile, the 8~cm$^{-3}$ model predicts the shape very well but is about 50\% more luminous than the observations. These small discrepancies are likely due to our constant density and spherical symmetry simplifying assumptions. While there are small differences between the models and observed emission line profiles, overall {\tt Cloudy} predicts the emission from ionized gas very well with simple models in a narrow density range, except for a few transitions that we will discuss in detail. 

In addition to the projected surface brightness profiles from the {\tt Cloudy} models, we also report the predicted total intensities in the {\tt Cloudy} models in \autoref{tab:lines} for the $4 \ \cm$ and $10 \ \cm$ models. Overall the radial profiles of the surface brightness provide a much more precise way of differentiating between the constant density {\tt Cloudy} models than examining the integrated intensity. For example, the integrated intensity of [\ion{Ne}{3}] varies between the $4 \ \cm$ and $10 \ \cm$ models by 28.33 to 29.38 $\rm W \ m^{-2}$, a $<1\%$ difference. In contrast, the radial profile of the $4 \ \cm$ model peaks 10~pc after the $10 \ \cm$ does, clearly showing it does not match the spatial distribution of the [\ion{Ne}{3}] emission. Thus a comparison between the spatial resolved radial profiles provides much more information than using just integrated intensities. 

To produce both the {\tt Cloudy} models and the measured radial profiles we assume that N76 is spherically symmetric with AB7 in the center of the nebula. This is approximately true, but not precisely correct. \autoref{fig:linemaps} shows that AB7 appears offset in the eastern direction. Additionally, there is an overall trend for the ionized gas lines to be much brighter on the eastern edge of the nebula in comparison to some of the neutral gas lines (\cii\ especially) that have brighter emission on the western edge. We explore the effects of the asymmetries on our results in Appendix \ref{sec:app}. Images of the \htwo\ S(1) transition and ancillary ALMA CO data (\autoref{fig:H2}) confirm that there is a molecular cloud complex on the western side of N76. Therefore, there are likely internal density gradients as well as a wall of dense gas on the western side of the nebula breaking the symmetry assumption. For example, because AB7 is slightly offset from the center, the radial averaging produces flatter slopes with respect to the models. We have attempted to account for some of this offset by showing the profiles as a strip instead of radially in \autoref{sec:app}. These show the structure in the different sides of the nebula and how a particular density is a better predictor for one side compared to the other.  

We also report the ionization structure predicted by {\tt Cloudy} of each element in \autoref{fig:ion} to complement the projected profiles of surface brightness. The panels show the ionization state of the relevant elements as a function of the physical distance from AB7. In addition to the ions that are studied in this paper, we also plot the ionization structure of helium. As the most abundant element after hydrogen, helium can have a significant impact on the ionization equilibrium of \hii\ regions. This impact is most pronounced in \hii\ regions illuminated by hard radiation fields, due to the fact that their central sources produce a lot of photons capable of ionizing and doubly ionizing helium: the ionization potentials of $\rm He^{0}$ and $\rm He^{+}$ are 24.6 eV and 54.4 eV respectively. Ionization cross-sections are sharply peaked at the ionization energy: the cross-section of $\rm He^{0}$ to 24.6~eV photons is ten times higher than that of $\rm H^{0}$, therefore these photons will preferentially ionize helium over hydrogen. For sources with a hard enough spectrum, characterized by the ratio of the rate of production of He-ionizing photons ($Q_1$) to H-ionizing photons ($Q_0$), this can lead to the $\rm He^{+}$ region extending slightly past the $\rm H^{+}$ region. In fact, $Q_1/Q_0\sim0.15$ is needed for the \hii\ and \heii\ regions to be equivalent sizes \citep{Draine2011}, while for N76 the $Q_1/Q_0\sim0.55$ \newchanges{calculated from the {\tt Cloudy} models} matches the slightly larger HeII emitting region \newchanges{seen in \citet{Naze2003}}. A similar effect is seen in the ionization structure of neon, where the [\ion{Ne}{2}] emission extends slightly past the hydrogen ionization front.

% \subsection{Emission lines that are inconsistent with photoionization models}
% \begin{itemize}
%     \item Discussion of the disagreement between the high ionization lines ([\ion{O}{4}] and [\ion{Ne}{5}]) with the photoionization modeling. How important are shocks? Or is the ionizing SED not well known?
%     \item The neutral dominated lines: \cii, [\ion{O}{1}], [\ion{Si}{2}] and the complexities of modeling a PDR. Depletion also plays a role, especially in [\ion{Si}{2}] 
% \end{itemize}

\begin{figure*}
    \centering
    \includegraphics[width=\textwidth]{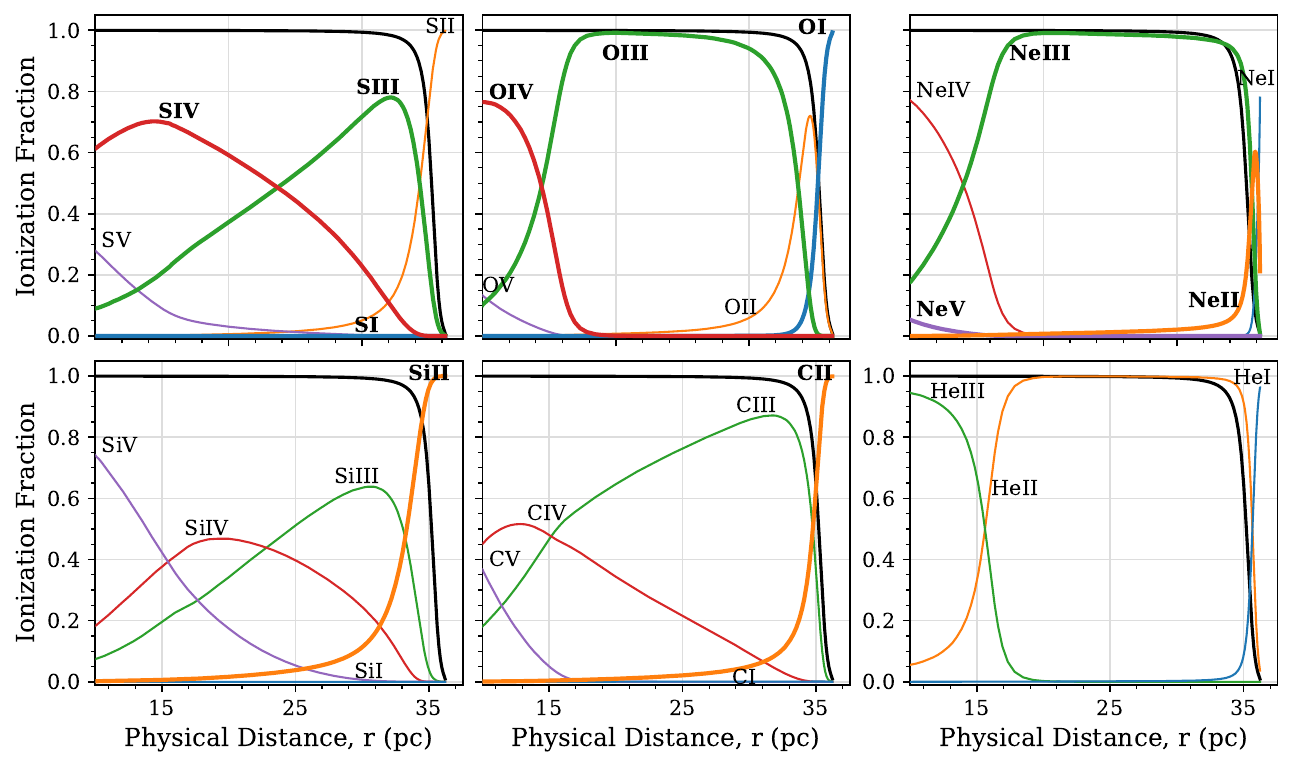}
    \caption{The ionization structure of sulphur, oxygen, neon, silicon, carbon, and helium from the 10 \cm\ constant density {\tt Cloudy} model beginning at the photoionized region (r = 10~pc). Ions that are presented in this paper are bold, and we also show the structure of ionized hydrogen in black. The various ionization zones can be seen throughout these elements in physical distance, similar to the projected distance seen in \autoref{fig:linemaps}.}
    \label{fig:ion}
\end{figure*}

\subsection{The high ionization lines: [\ion{Ne}{5}] and [\ion{O}{4}]}\label{sec:high_ion}
The high ionization lines, [\ion{Ne}{5}] and [\ion{O}{4}], with ionization potentials of 97 eV and 55 eV respectively, can often be difficult to reproduce through just \changes{photoionization from O and B stars}. Models of hot, low metallicity WR stars around ages of 3-4 Myr like the WR in AB7, however, contain enough high energy photons to ionize Ne$^{+4}$ and O$^{+3}$ and produce appreciable [\ion{Ne}{5}] 24 \micron\ and [\ion{O}{4}] 25 \micron\ emission \citep{Schaerer1998, Schaerer1999, Crowther1999}. The N76 photoionization models presented here are able to accurately predict both the flux of the [\ion{Ne}{5}] and [\ion{O}{4}] lines (see \autoref{tab:lines}) as well as their spatial distribution (see \autoref{fig:radprof_den}). The [\ion{Ne}{5}] is slightly over-predicted by the {\tt Cloudy} models by a factor of two, which can be explained by density variation \changes{(either through clumping or a lower density in the high ionization zone)} or uncertainty in the PoWR model \changes{(see below)}. 

We note that the predicted flux of the [\ion{Ne}{5}] and [\ion{O}{4}] lines are highly dependent on small changes to the PoWR model that impact the high energy (55+ eV) tail of the SED. A {\tt Cloudy} model using a PoWR SED with the same input parameters but increasing the hydrogen fraction from $X_{H} \sim 0.15$ to $X_{H} \sim 0.2$ produced no [\ion{Ne}{5}] emission and a fraction ($<10\%$) of the observed [\ion{O}{4}] emission. Additionally, the O star companion to the WR in AB7 contains a small X-ray contribution to the SED, implied by the presence of OVI in its spectrum, but contributes a negligible ($<5\%$) amount to the [\ion{O}{4}] and [\ion{Ne}{5}] flux \citep[see][for more details]{Shenar2016}. 

Further, \citet{Naze2003} report narrow-band optical imaging of N76 and find bright HeII $4686 \ \rm \AA$ emission that is inconsistent with the hottest WR models at the time. The {\tt Cloudy} models presented here, however, are able to reproduce the reported HeII emission within a factor of two. The spatial distribution of the HeII emission \citep[see Figure 4 in][]{Naze2003} is very similar to the [\ion{O}{4}] emission, likely due to their similar ionization potentials.

\subsection{The wind-blown cavity}
\label{sec:wind}

The {\tt Cloudy} photoionization models require a 10~pc cavity to reproduce the observed spatial distribution of the ionized gas lines. This origin of this cavity is from the stellar winds of the WR and O star in AB7 interacting, shocking, and heating the surrounding ISM, creating a stellar wind-blown bubble (SWB). The resulting nebula consists of an onion like structure \citep[see Figure 1 in][]{Weaver1977, Freyer2003} with four layers, including (1) an innermost, free-flowing supersonic wind that travels into (2) a hot ($\sim 10^6 - 10^8$ K), shocked gas region that will expand into (3) the photoionized shell of swept up interstellar gas, and finally into (4) the ambient interstellar medium. The hot, shocked material in shell 2 is typically at much higher pressures than shell 3, and will expand as a function of time. Assuming a uniform density spherical shell with equal pressure throughout and that the thermal energy contained within shell 2 is much higher than the kinetic energy, \citet{Weaver1977} derived a relationship between the radius of hot, shocked gas bubble (shell 2) as a function of time, such that

\begin{equation}
\label{eq:rs2}
    r_{s2}(t) = \left( \frac{125}{154 \pi} \right)^{1/5} L_w^{1/5} \rho_0^{-1/5} t^{3/5} \ \rm{cm},
\end{equation}
\noindent where $\rho_0 = n_0 m_H$ is the density of the ambient medium in $\rm g \ cm^{-3}$, $L_w = \frac{1}{2} \dot{M_w} v_w^2$ is the stellar wind luminosity in $\rm g \ cm^{2} \ s^{-3}$ with $\dot{M_w}$ as the mass loss rate and $v_w$ as the terminal velocity of the stellar wind, and $t$ is the time of expansion in seconds. We calculate the size of the bubble blown by the WR star in AB7 through the values characterized by \citet{Shenar2016}, $\dot{M} = 10^{-5} \ \mathrm{\msun \ yr^{-1}}$, $v_w = 1.7 \times 10^{3} \ \mathrm{km \ s^{-1}}$, and assume $n_0 = 10 \ \mathrm{cm^{-3}}$. Using these parameters, a 10~pc radius bubble is created after 0.2 Myr of free expansion. This is consistent with the evolutionary tracks studied in \citet{Shenar2016}, where they find an age of 3.4~Myr for the AB7 system and estimate the WR star has been in the WR stage for $\sim$0.1-0.2~Myr. 

Another contribution to the cavity in N76 may be through the colliding stellar winds from the WR and O star. There is ample evidence of shocks originating from colliding wind systems, such as the winds produced by the WN4 and O6 star in AB7 \citep[e.g.,][]{Stevens1992, Tuthill1999, Parkin2008}. However, the spatial scale of the colliding wind zone is comparably insignificant to size of the photoionized cavity, corresponding to $\sim 10^{-6}$~pc for AB7. Further, \textit{Chandra} observations of AB7 show \newchanges{X-ray} emitting gas attributed to shocks from the colliding winds, but the region is only $\sim$3.5~pc in radius \citep{Guerrero2008}. Thus the SWB structure described in \citet{Weaver1977} likely creates the implied cavity in the presented photoionization models. 

\changes{It is possible that shocks from either the colliding winds of the AB7 system or the stellar wind blown bubble could contribute additional sources of [\ion{Ne}{5}] and [\ion{O}{4}] in N76. We investigate this} by calculating the velocity and postshock temperature needed to ionize Ne$^{+3}$ into Ne$^{+4}$. We estimate a temperature of $\rm T \sim E_{ion}/k_B \sim 10^6$~K by taking the ionization energy of [\ion{Ne}{5}]. Using the relationship between the postshock temperature \citep[equation 36.28 of ][]{Draine2011} and the shock velocity, we find the minimum velocity for the shock is $\rm v \approx 290$ \kms. \citet{Toala2017} study the Wolf-Rayet Nebula NGC 3199 in the Milky Way and find diffuse X-ray emission that corresponds to a plasma temperature of $\rm T \approx 1.2 \times 10^{6} \ K$, very similar to the temperature required to produce the [\ion{Ne}{5}] emission observed in N76. Further, models of stellar wind blown bubble show velocities of $\sim200-400$~\kms, enough to ionize Ne$^{+3}$ into  Ne$^{+4}$ \citep{Freyer2003, Freyer2006, Lancaster2021}. Since the {\tt Cloudy} photoionization models already slightly overpredict the [\ion{Ne}{5}] emission, it is unlikely there is this additional contribution from shocks. The cavity may contain ions of neon at much high energy states not detected by our current dataset.

Fast, radiative shocks with velocities $\sim$300-500 \kms\ can explain optical observations of [\ion{Ne}{5}] and HeII in low metallicity, blue compact dwarf galaxies \citep[e.g.,][]{Thuan2005, Izotov2012, Izotov2021}. However, the origin of these shocks is typically attributed to supernovae (SN), not stellar winds. Indeed, the supernova remnant E0102-72 that serendipitously is in the N76 field is a bright [\ion{Ne}{5}] source (see \autoref{fig:linemaps}), suggesting shocks from SN can produce appreciable [\ion{Ne}{5}] emission. \changes{The SNR is too far away from N76 to contribute to the [\ion{Ne}{5}] emission associated with the nebula (see \autoref{fig:linemaps}). Therefore, the [\ion{Ne}{5}] within the N76 nebula is most likely produced through photoionization. We note that if N76 was placed at the larger distances of the blue compact dwarf galaxies studied in \citet{Izotov2021}, it would be difficult to differentiate between [\ion{Ne}{5}] produced through the Wolf-Rayet nebula and the SNR due to the lower physical resolutions.}

% This is then compared to the estimated temperature given by \citet{Weaver1977}:
% \begin{equation}
% \label{eq:T_s2}
%     T_{s2} = 2.07 \times 10^{6} L_{36}^{8/35} n_{0}^{2/35} t_{6}^{-6/35} (1 - \eta)^{2/5}
% \end{equation}

%\subsubsection{SED for the Wolf-Rayet star harder than predicted?}
%It is also possible to produce the high ionization lines, [\ion{O}{4}] and [\ion{Ne}{5}], through just photoionization if the models of the energy output of the WR star are harder than the PoWR model predicts. The ionizing spectrum of WR stars greater than 13.6 eV is uncertain (citation) due to the photon absorption in the HII region. If the luminosity of the WR in AB7 stays constant, but the photon energy is a factor of two harder (the SED produces photons up to $\sim$100 eV instead of $\sim$50 eV), then there are enough ionizing photons to produce the observed intensity of [\ion{O}{4}] and [\ion{Ne}{5}]. This model requires a $\sim$10~pc wind-blown bubble in order for the emission to peak at the right position in the HII region. 

\subsection{The neutral \newchanges{gas} dominated lines: [\ion{O}{1}], [\ion{C}{2}], and [\ion{Si}{2}]} \label{sec:result_neut}

We define the neutral-dominated lines as ions or atoms that can come from neutral gas, i.e. those ions that have an ionization potential less than that of hydrogen. Most of these neutral-dominated emission lines are under-predicted by the {\tt Cloudy} models \changes{because} the {\tt Cloudy} models were designed to model the ionized gas and do not probe farther than the ionization front after the transition from ionized to neutral material. Further, \autoref{fig:ion} shows the ionization structure of carbon and oxygen, suggesting that the radiation field of AB7 is too hard for $\rm O^{0}$ and $\rm C^{+}$ to exist in the interior of the \hii\ region where oxygen and carbon are in higher ionization states. Thus the bulk of emission from [\ion{O}{1}] and [\ion{C}{2}] comes from the neutral phases; future photo-dissociation region (PDR) modeling is needed to understand the nature of these neutral lines, which is beyond the scope of this study. 

The fine-structure transition of silicon, [\ion{Si}{2}], however, is reasonably well predicted by the {\tt Cloudy} models. With an ionization potential of 8.15 eV for $\rm Si^{0}$, [\ion{Si}{2}] can come from the neutral gas. However, {\tt Cloudy} modeling of the \hii\ region reproduces both the total intensity and the surface brightness distribution of the [\ion{Si}{2}] line (see \autoref{tab:lines} and \autoref{fig:radprof_den}), unlike for the other neutral gas tracers. \autoref{fig:ion} shows that the modeled abundance of Si$^+$ grows faster toward the edge of the \hii\ region than that of C$^+$, \changes{which is consistent with the observed radial profiles presented in \autoref{fig:radprof_den}. Thus the majority of the [\ion{Si}{2}] emission is associated with the edge of the \hii\ region, \newchanges{around} the transition between ionized and atomic gas. In \autoref{sec:disc-ion} we calculate the percentage [\ion{Si}{2}] comes from each phase.}

%The key difference is that the critical density of [\ion{Si}{2}] for collisions with electrons is large ($\rm n_{crit} = 2 \times 10^{3}$~cm$^{-3}$), compared to the critical density of [\ion{C}{2}] ($\rm n_{crit} = 5 \times 10^{1}$~cm$^{-3}$). 

%Further, the excitation of [\ion{Si}{2}] requires higher temperatures than that of [\ion{O}{1}]. Thus [\ion{Si}{2}] emission will be found in dense, hot areas around \hii\ regions. \autoref{fig:linemaps} shows that the [\ion{Si}{2}] emission is a skin around the \hii\ region, marking the transition between the ionized gas and the PDR at the edge of the ionization front. The other neutral lines, [\ion{C}{2}] and [\ion{O}{1}], may be in the lower-excitation portions of the PDR where there is a high enough temperature and density to excite [\ion{O}{1}] and [\ion{C}{2}] but not [\ion{Si}{2}]. 

Why are there dominant PDR contributions for [\ion{C}{2}] and [\ion{O}{1}] but not [\ion{Si}{2}]? The [\ion{Si}{2}] 34 \micron\ fine-structure transition is at shorter wavelengths, and its excitation requires higher temperatures than [\ion{O}{1}] 63 \micron\ or [\ion{C}{2}] 158 \micron . Further, the [\ion{C}{2}] and [\ion{O}{1}] lines are also major coolants in the Warm and Cold Neutral Medium \citep{Wolfire1995, Wolfire2003} and contributions from that can increase the total intensity of these lines by peering through the atomic gas in projection towards the SMC. In contrast, \newchanges{models presented in \citet{Wolfire1995, Wolfire2003} show the cooling from [\ion{Si}{2}] in the atomic gas} is two orders of magnitude lower than that from the [\ion{C}{2}] and [\ion{O}{1}] lines, which leads to a much fainter \newchanges{atomic gas} contribution for the [\ion{Si}{2}] emission. The intensity of the [\ion{Si}{2}] emission is also highly dependent on the silicon gas abundance. Silicon is depleted heavily into dust, and in order to determine its gas phase abundance we used depletion studies in the SMC \citep{Tchernyshyov2015, Jenkins2017}. The dust depletion of gas phase silicon with respect to the photospheric abundances in the SMC \citep{Russell1992} has a large range, factors from 1.25 to 0.05 (0.1 to -1.3 dex, our preferred abundance corresponds to 0.4, or -0.4 dex). \newchanges{The assumed silicon abundance matches well the observed radial profile in in \autoref{fig:radprof_den}. A lower silicon abundance leads to a [\ion{Si}{2}] radial profile that does not match the overall shape and would suggest a larger fraction of [\ion{Si}{2}] comes from the PDR.} While we assume some silicon depletion in the ionized gas, our results suggest that even more silicon is depleted in the neutral phase, especially when compared to carbon or oxygen, \changes{which will decrease the available gas-phase silicon in order to produce the [\ion{Si}{2}] emission}. These combined effects likely explain the much smaller contribution of PDRs toward the observed [\ion{Si}{2}] emission.
% is likely locked in dust in neutral and/or quiescent regions.

We present images of the other neutral gas tracers in N76 in \autoref{fig:H2}. The \spitzer\ IRS covers the quadrupole transitions of molecular hydrogen, \htwo. Our PAHFIT pipeline simultaneously fits these lines and we show the brightest \htwo\ line, S(1) at 17 $\mu$m, in \autoref{fig:H2}. This line has $E/k \sim 1000$~K, traces \htwo\ at $T>300$~K, and its presence indicates molecular material that is heated by a strong FUV field or through shocks from the central WR star. We compare the \htwo\ S(1) and [\ion{Si}{2}] lines to the ACA CO(2-1) data presented in \citet{Tokuda2021}. Interestingly, these neutral gas species are not always spatially coincident. As discussed above, the [\ion{Si}{2}] emission originates mostly \newchanges{around the ionization front of the \hii\ region} and therefore traces the the outer shell of N76. The \htwo\ and CO emission is brightest where AB7 illuminates the molecular ridge region that is north and north-west of N76.

\begin{figure}
    \centering
    \includegraphics[width=\columnwidth]{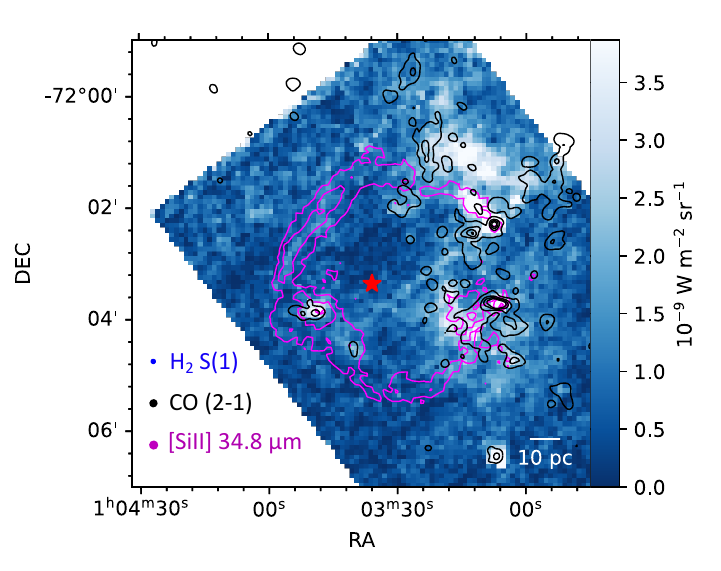}
    \caption{The neutral gas tracers in N76, with the \htwo\ S(1) quadrupole transition in the blue color scale, CO contours from \citet{Tokuda2021} in black (in levels of 1, 6, 11, 16 Jy/beam), and the [\ion{Si}{2}] emission in magenta contours (in levels of 1, 1.5, and 3 $\rm \times \  10^{-8} \  W \ m^{-2} \ sr^{-1}$). The beam/PSF size of each map is in the bottom left corner. The [\ion{Si}{2}] emission primarily traces the outer shell of N76, \changes{suggesting [\ion{Si}{2}] originates mostly from the transition between the ionized gas and the PDR}. The CO and \htwo\ emission is brightest when AB7 illuminates the molecular ridge north and north-west of N76.}
    \label{fig:H2}
\end{figure}

% this is a non conclusion 
% SiII traces the shell of the HII region even when there is no molecular gas. 

\subsection{Production of [\ion{Ne}{2}] in N76}
\label{sec:NeII}

The {\tt Cloudy} models predict a modest amount of [\ion{Ne}{2}], about 20\% of the observed [\ion{Ne}{2}]. At an ionization potential of 21.56 eV, [\ion{Ne}{2}] traces lower energy photons. \autoref{fig:ion} shows how the ionization state of Neon transitions to $\rm Ne^{+}$ at the very edge of N76 \changes{at a very small radius range compared to the higher ionization states and rapidly transitions into neutral Neon. Therefore, the total volume of the predicted [\ion{Ne}{2}] emission from the {\tt Cloudy} models is quite small, leading to a very weak predicted [\ion{Ne}{2}] line.} The excess of observed [\ion{Ne}{2}] emission could be associated with other ionization sources besides AB7, such as smaller \hii\ regions. It is also possible that [\ion{Ne}{2}] traces the low density, diffuse ionized gas at densities $n_e \lesssim 0.1 \ \rm cm^{-3}$. In the nearby dwarf galaxy IC~10, \citet{Polles2019} also find that their models underpredict the [\ion{Ne}{2}] observations by a factor between three and four, suggesting that [\ion{Ne}{2}] may be a tracer of the diffuse ionized gas in low metallicity galaxies.

\subsection{How changes in the model affect the inferred conditions}
% \begin{itemize}
%     \item Comparison of constant density vs. constant pressure models
%     \item Discussion of stopping conditions in {\tt Cloudy}
% \end{itemize}

Besides the density, for which we produce a grid of models, a variety of other {\tt Cloudy} inputs can affect the final results of the modeling. The {\tt Cloudy} models are most sensitive to the input luminosity and SED. 

For example, decreasing the luminosity of the WR by 50\% leads to a 56\% reduction in the total intensity of the spectral lines at a constant density of  $n_H = 10 \ \cm$. Naively, this reduction in the total intensity provides a slightly better fit for some lines, but it is at the cost of producing emission over a region with a smaller radius and a spectral line radial profile that does not match the data. Thus the strength of comparing the {\tt Cloudy} models to the spatially resolved data shines through; using the {\em a priori} values for the spectrum and luminosity provide an overall better fit than if one were to directly match the total intensities. 

We also experimented with using different stopping conditions, abundances, and the presence of dust. All of these parameters have a negligible impact on the overall results of the model, except for the abundances which we fixed in the manner described in $\S$\ref{sec:parameters} and \autoref{tab:abun}.  

\subsection{Morphology of emission lines}
%While the radial profiles accurately trace the ionization structure of N76, our resolved images reveal the structure of this \hii\ region, including a variety of features. 
Below we discuss some features that are apparent in our images of the N76 region shown in \autoref{fig:linemaps}.

\textit{Sharp and bright eastern edge.} In most of the ionized gas emission lines, there is a brightness gradient with a northeastern sharp edge compared to a fainter and more diffuse southwestern edge. This is especially prominent in the lower ionization lines such as [\ion{S}{3}]. We test how the brightness gradient impacts the {\tt Cloudy} modeling in \autoref{sec:app}. Modeling each individual side suggests a density difference of order $\pm2$ \cm\ from the best density in the radially averaged models, or a typical density difference between each side of approximately $20\%-30\%$. 
%The radial profiles create an average between the lower and higher density sides, but there is density structure in this nebula. 
% may be related to an interaction with the powerful winds sweeping towards the nebula, building up material.

\textit{Potential Spiral structure.} In some of the high ionization lines, particularly [\ion{O}{4}] and [\ion{S}{4}], there appears to be a \changes{``pinwheel''} spiral structure emanating from AB7. \changes{The spiral has a single arm beginning at the location of AB7 and coils eastward then westward, ending about 10~pc away from AB7. This structure is also present in narrow-band optical line imaging of HeII $4686 \ \rm \AA$ from \citet{Naze2003}}. Spirals in WR-O star binaries have been seen before in dust emission \citep[e.g.,][]{Tuthill1999,Lau2022} caused by the interacting stellar winds and the orbits of the binary. Spiral features are typically detected in the dust continuum of carbon-rich WC stars, but a spiral structure was reported in a nitrogen-rich star similar to the Wolf-Rayet in AB7 in radio continuum likely due to synchrotron emission \citep{Rodriguez2020}. This suggests that although the presence of dust helps illuminate a spiral pattern, it is intrinsic to binary systems and also present in other tracers. The spiral features documented in the literature, however, are on the scale of hundreds or thousands of AUs rather than the $\sim$10~pc feature we see in N76. The wind-collision zone where a spiral structure is expected for AB7 is $\sim$2 AU in size, and therefore unresolved in our data. Nevertheless, it is possible that the high ionization lines of N76 show an imprint of the spiral motions from past stellar wind collisions in the central region, which then expand with the wind. Alternatively, the suggestive spiral structure may be coincidental and due to density gradients throughout the photoionized gas. 

\textit{Dense knot.} In the southeastern edge of N76 there is a bright knot of emission seen across the \spitzer\ IRS low ionization emission lines (and especially prominent in the [\ion{Si}{2}] line). Previous work studying N76 suggest that this dense region is an independent compact \hii\ region occasionally referred to as a ``high excitation blob'' \citep{Heydari2001, Naze2003}. These are very dense, small regions that are associated with young massive stars beginning to leave their natal molecular cloud. \autoref{fig:H2} shows the highly excited molecular gas in this blob, both in the \htwo\ quadrupole transitions as high as S(7) and a small clump of CO (2-1) \citep{Tokuda2021}. There are two possible ionizing sources for this compact region, including a Be star \citep{Wisniewski2006} and a candidate Young Stellar Object \citep{Bolatto2007}. It is unclear whether either of these stars can power the ionization seen in this clump. Alternatively, it is possible that this feature is simply a dense blob of gas irradiated by AB7 and shocked by the \hii\ region expansion. 

% \textit{Arm.} In most of the ionized gas emission lines from \spitzer\ IRS, there is an arm forking from the central region of N76. This feature may be due to material outflowing from the central nebula or a density gradient in the emission. 
%It is also possible that the arm is a pillar-like structure, it ends near the dense blob which contains highly excited molecular gas seen in both \htwo\ S(1) and CO (2-1) (see \autoref{fig:H2}. 
% too vague of a section without much evidence backing up-- arm could also be part of the spiral structure. Remove unless requested. 

\subsection{Tentative [\ion{S}{1}] detection in the center of N76}
\label{sec:SI}

One peculiar detection from \spitzer\ IRS is a spectral line at 25.25 \micron\ in the center of N76 that we tentatively identify as [\ion{S}{1}]. The spectrum of this line can be seen in \autoref{fig:spec} and the image in \autoref{fig:linemaps}. Our fitting procedure from PAHFIT identifies a spectral line between 25.2 and 25.3 \micron\ throughout the central region of N76. The neutral sulphur line is directly in the middle of this range, at 25.245 \micron. Using the NIST atomic spectral database, the nearest possible spectral lines correspond to Mg~II at 25.05~\micron\ and Si~I at 25.38~\micron. The redshift of these wavelengths due to the systemic velocity of the SMC is $0.01$~\micron. The average uncertainty PAHFIT reports on this spectral line wherever there are detections is 0.017~\micron, which corresponds to $\sim10\%$ of the wavelength step between spectral pixels in this region of the LL1 map (0.178~\micron). 

The existence of neutral sulphur (or for that matter any line corresponding to ions found in neutral gas, such as Mg~II and Si~I) in the center of such a high ionization \hii\ region is a mystery since neutral sulphur should be ionized quickly nearby such a hot and luminous star. \changes{Indeed, {\tt Cloudy} predicts a total of $0.0005 \ \times 10^{-15} \rm \ W \ m^{-2}$ for the total [\ion{S}{1}] intensity, which is almost four orders of magnitude fainter than the observed total flux of $1.04 \times \ 10^{-15} \rm  \ W \ m^{-2}$. The radial profile {\tt Cloudy} predictions are also many orders of magnitude fainter than the observed flux, predicting an average of $\sim 10^{-13} \ \rm W \ m^{-2} \ sr^{-1}$ compared to the observed $\sim 10^{-9} \ \rm W \ m^{-2} \ sr^{-1}$ radial profile.} If the line is confirmed as [\ion{S}{1}], then it is the first detection of such a feature in the center of the highly irradiated environment of a WR nebula. Previous observations show [\ion{S}{1}] resides in dense, neutral gas and can be produced through fast, dissociative J-shocks \citep{Haas1991, Rosenthal2000}. These observations, however, consistently show the presence of other neutral lines, such as \htwo\ and [\ion{O}{1}]. In N76, there is no corresponding neutral emission where we observe [\ion{S}{1}], making the production of this line a mystery. We explore possible explanations for this emission in $\S$\ref{sec:SI_ideas}.
 
% It is perhaps possible that we have misidentified this line, but [\ion{S}{1}] is our best candidate.

\section{Discussion}
\label{sec:discussion}

\subsection{Ionized gas contributions to the neutral lines: [\ion{Si}{2}], [\ion{C}{2}], and [\ion{O}{1}]}
\label{sec:disc-ion}

The ions and atoms that can exist in the neutral gas, $\rm Si^{+}$ and $\rm C^{+}$, have ionization potentials less than that of hydrogen, but their next ionization stage is above or close to 13.6~eV. The emission from [\ion{Si}{2}] and [\ion{C}{2}] can therefore be present in both the neutral and ionized gas phases. \newchanges{For [\ion{O}{1}], the ionization potential of oxygen (13.618 eV) is very close to that of hydrogen and facilitates a very efficient charge-exchange reaction, ensuring that there is very little [\ion{O}{1}] emission associated with the ionized gas.} Although this work is focused on modeling only the ionized gas, understanding the proportion the ionized gas contributes to the overall intensity of the neutral-dominated lines is of interest for using these lines as neutral gas tracers. Our {\tt Cloudy} simulations were designed to end \changes{at} the ionization front, the position where the ionized gas transitions into fully neutral gas. \changes{To match the method in \citet{Cormier2019}}, we compute the contribution the neutral gas has on [\ion{Si}{2}] and [\ion{C}{2}] by defining a more conservative ``end'' of the ionized gas region as the point in which the electron fraction and \newchanges{proton} fraction are equal to 0.5. We use this value to define the end of the ionized gas region.

As described in $\S$\ref{sec:result_neut}, the {\tt Cloudy} models underpredict the emission from [\ion{O}{1}] and [\ion{C}{2}] that is observed in the direction of the nebula, because the neutral gas regions that surround the ionized gas but are not included in our {\tt Cloudy} models contribute the bulk of the emission in these lines. In contrast, the [\ion{Si}{2}] intensity is well matched to the predicted emission from ionized gas in the {\tt Cloudy} models, suggesting a substantial amount of [\ion{Si}{2}] is produced \changes{at the edge of the} \hii\ region itself and therefore not outside the \hii\ region in the PDR. We quantify this by calculating the fraction of the intensity originating from the ionized gas for these lines in our models \changes{by defining the end of the ionized gas region as the point where the electron and neutral fraction are equal to 0.5}. We find that for N76 91\% of the [\ion{Si}{2}] emission originates from the ionized gas while the [\ion{C}{2}] has an ionized gas contribution of 3\%. \citet{Cormier2019} use their {\tt Cloudy} models of the Dwarf Galaxy Survey (DGS) to perform a similar calculation and report that $\lesssim$10\% of the [\ion{C}{2}] and $\lesssim$40\% of [\ion{Si}{2}] emission originates from the \hii\ regions. Although we see a much higher \hii\ region contribution to the [\ion{Si}{2}] emission in N76 than in the galaxy-wide integrated measurements in the DGS, keep in mind that N76 is a Wolf-Rayet nebula and thus not representative of most \hii\ regions. 

%This suggests that substantial fractions of [\ion{Si}{2}] emission may exist in the warm ionized medium that pervades on galaxy scales. Our measurements focusing only on a single \hii\ region are not able to trace any broad scale [\ion{Si}{2}] emission. 

\subsection{Production of [\ion{S}{1}] in the center of N76}
\label{sec:SI_ideas}
We discussed in $\S$\ref{sec:SI} the surprising identification of a line in the central regions of N76 as neutral sulphur. Below we go over some possible ideas about the origin of this transition.

\textit{Velocity shifted [\ion{O}{4}] emission}: The nearest emission line in wavelength to our possible [\ion{S}{1}] emission that is bright and clearly identified in N76 is [\ion{O}{4}] at 25.93 \micron. We considered whether the line we identify as [\ion{S}{1}] could be high velocity [\ion{O}{4}] emission. The velocity required to shift [\ion{O}{4}] to the observed wavelength of 25.25 \micron\ is 7,600 \kms. While the WR and O stars in AB7 produce stellar winds, the speed of these winds is below 2,000 \kms\ \citep{Shenar2016}. The X-ray emission observed in N76, resulting from the colliding WR and O star winds, also implies a wind speed of 2,000 \kms\ \citep{Guerrero2008}. Thus it is extremely unlikely that the emission is [\ion{O}{4}] shifted to 25.25 \micron. Further, the wind would have to be asymmetric, since we do not see a corresponding redshifted component. Moreover, instead of discrete components we would expect a symmetric, broadened [\ion{O}{4}] profile because the emission should be optically thin. 

\textit{Dust destruction through \newchanges{shocks from} stellar winds}: The main mystery of the origin of [\ion{S}{1}] at the center of this nebula is how sulphur can remain in a neutral state where ionized species caused by $>$55~eV photons are also observed. It is possible that the winds and radiation from AB7 can \newchanges{shock} the surrounding dust, destroy dust grains, and move their constituent atoms, including neutral sulphur, into the gas phase. We then observe the [\ion{S}{1}] emission before neutral sulphur is ionized into higher states. \newchanges{Most previous detections of [\ion{S}{1}] are associated with shocks \citep[e.g.,][]{Haas1991, Rosenthal2000, Neufeld2009, Simpson2018} and shocks provide a mechanism to both release neutral sulphur material from the dust grains into the gas phase and to excite [\ion{S}{1}] emission.}

We evaluate whether this is possible by performing a back-of-the-envelope calculation. First, we estimate the timescale it would take to ionize sulphur in the hard radiation field of AB7. We integrate the ionizing spectrum over the ionization potential of sulphur (10.36 eV) to find the rate of sulphur-ionizing photons at a given distance of the [\ion{S}{1}] emitting region. We calculate a timescale of 1.7 years to ionize sulphur after assuming a cross-section of $\rm 3 \times 10^{-18} \  cm^{-3}$ \citep{Draine2011} and a radius of 10~pc. We next calculate the mass of neutral sulphur that must be replenished within this time. From the observed [\ion{S}{1}] intensity we find the column density by assuming collisional excitation and use the equation given in \citet{crawford1985}, a critical density for collisions with $\rm e^{-}$ of $\rm n = 1.55 \times 10^{5} \ cm^{-3}$ \citep{Draine2011}, a spontaneous decay rate of $\rm A = 1.40 \times 10^{-3} \ s^{-1}$ \citep{Froese_Fischer2006}, a temperature of $\rm T = 1.4 \times 10^{4} \ K$, an abundance of $3.89 \times 10^{-6}$ \citep{Russell1992}, and a density of $\rm n = 40 \ cm^{-3}$. The density is chosen by assuming a strong shock wave will have four times the density of the pre-shocked medium of 10 \cm \citep[e.g.,][]{Draine2011}. We calculate a column density of $\rm N_{S^{0}} = 4.7 \times 10^{14} \ cm^{-2}$  and consequently a mass of $\rm M_{S^{0}} = 3.7 \times 10^{-2} \ \msun$ over a 10~pc radius. Thus our estimate suggests that $\rm 3.7 \times 10^{-2} \ \msun$ of sulphur will need to be lifted off of destroyed dust grains every 1.7 years. Given the low abundance of sulphur relative to other elements that make the bulk of the dust mass (C, Si, O), this represents a very large dust mass destroyed every year, and it seems impossible this reservoir would exist over the life of the nebula (or be in some way replenished by the central source mass loss rate). \newchanges{Further, there is no signature of other neutral species besides [\ion{S}{1}] near the center of the nebula to support the scenario that dust destruction from shocks could produce the neutral sulphur.}

Further, we would also expect that the other neutral species present in dust are also at the central part of nebula, which is not the case.
%Given that the mass loss rates of AB7 are $\rm log \dot{M} = -5 \ \msun \ yr^{-1}$, it is extremely unlikely that there is enough dust to replenish the observed [\ion{S}{1}] emission. 

\newchanges{We note that if the [\ion{S}{1}] emission is produced through dust destruction due to shocks from stellar winds, there would likely be a substantial amount of [\ion{Fe}{2}] emission produced from the same mechanism since iron is heavily depleted in dust. The 25.9 \micron\ blend of [\ion{O}{4}] 25.89 \micron\ emission and [\ion{Fe}{2}] 25.99\micron\ described in \S\ref{sec:blend} may therefore have a substantial contribution from [\ion{Fe}{2}] 25.99 \micron . However, since the [\ion{Fe}{2}] 17.93 \micron\ line is not detected and we predict $<5\%$ of the 25.9 \micron\ line flux is from [\ion{Fe}{2}], we conclude that dust destruction scenario is unlikely.}

% stays neutral, cannot produce other neutral stuff we don't  see 
%\textit{Dust-shielded ``packets'' of neutral sulphur} The dust temperature reported from Cloudy models 1~pc away from AB7 ranges from 60 - 120 K. If we just consider the radiation from AB7 as the source of dust destruction, it is very likely that dust will survive where the [\ion{S}{1}] line is observed. Thus it is possible that tiny packets of dust can surround neutral sulphur, shielding sulphur from the incoming $>10.36$ eV photons that ionize it into higher states. Previous observations show [\ion{S}{1}] is primarily produced through fast, dissociative J-shocks \citep{Haas1991, Rosenthal2000}. These observations, however, consistently show the presence of other neutral lines, such as \htwo\ and [\ion{O}{1}], in addition to [\ion{S}{1}]. There appears to be no corresponding neutral emission in the proximity of where we observe [\ion{S}{1}], although there are upper limits. A careful treatment of shock modeling that goes beyond the scope of this paper is required to understand the production of [\ion{S}{1}] in N76. 

\changes{\textit{Artifact in Spitzer IRS Data}: The feature at 25.25 \micron\ has multiple detections through the spectral slit scans and is not a one time, random anomaly with the data. We instead explore two detector effects associated with \spitzer\ IRS data that can manifest as spurious spectral features and explain how they are unlikely to cause the possible [\ion{S}{1}] feature at 25.25 \micron . The IRS Handbook reports that spectral ``ghost'' features have appeared near extremely bright spectral lines in the short-high (SH) module near the 12.8 \micron\ [\ion{Ne}{2}] line \citep{Simpson2007, IRSHandbook}. However, it is extremely unlikely that the 25.25 \micron\ feature is a ghost of the 25.9 \micron\ [\ion{O}{4}] line as the IRS Handbook specifically reports that no similar features had been found for the long-low (LL) module that covers the 25.25 \micron\ line. The ghost features manifest in the SH module due to order mixing in the cross-disperser and there are no equivalent order-sorting filters for the LL module. Further, the ghost features that appear in the SH module are symmetric across the [\ion{Ne}{2}] line due to smearing of the flux on the plane of the slit and are also 0.5-1\% of the total flux of the line. The [\ion{S}{1}] 25.25 \micron\ feature observed in N76 is completely asymmetric, does not appear as a smeared feature on the slit plane, and is 5\% of the total [\ion{O}{4}] flux, therefore strengthening the position that it is unlikely for the [\ion{S}{1}] line to be a ghost feature of [\ion{O}{4}]. The other possible artifact we explore is spectral fringing, reported in the IRS handbook to be associated with the LL module where the 25.25 \micron\ line is found. Spectral fringing manifests as residual non-sinusoidal wiggles in the IRS spectrum and begins at $\sim$20 \micron\ \citep{IRSHandbook}. This fringing was more prevalent earlier in the \spitzer\ mission and has been seen in objects such as M82 and Cassiopeia A \citep{Beirao2008, Smith2009}. In these objects, the fringes are in a continuous pattern from 20 - 30 \micron , are 1-2\% of the flux, and are coincident with the continuum and correlated with its brightness. The 25.25 [\ion{S}{1}] \micron\ feature in N76 is therefore unlike these spectral fringes because only a single feature appears in the spectrum, not the continuous pattern seen in all other cases. It is also a factor of 2 or more brighter than the fringe features and the [\ion{S}{1}] line is not spatially correlated with the continuum. Therefore, we conclude that the 25.25 \micron\ line is likely not due to a \spitzer\ IRS artifact, but future observations (such as with JWST) will be need to confirm and characterize this curious line. }

\textit{Unidentified spectral line}: We used the NIST atomic spectral database to search for lines near the 20.0-20.5~\micron\ range and did not identify any possible candidates besides [\ion{S}{1}] (see discussion in $\S$\ref{sec:SI}). It is possible that the 25.25 \micron\ emission originates from a molecular compound, but a molecule would have similar difficulties surviving the intense radiation field of AB7. Instead of [\ion{S}{1}], however, this line could be an unidentified spectral line that corresponds to a very high ionization state not in the NIST catalog.

\subsection{Physical conditions in N76}
\label{sec:phy_con}
% \begin{itemize}
%     \item \underline{Figure:} show the temperature, cooling, etc profile from the Cloudy output
% \end{itemize}

\autoref{fig:phy} presents how the electron temperature and electron density vary for each constant \changes{hydrogen} density model as a function of physical distance from AB7. The electron density is slightly higher at the center of N76, due to the contribution from the $\rm He^{++}$ region that forms around very hot WR stars (see also sixth panel in \autoref{fig:ion}). In our constant \changes{hydrogen} density modeling $\rm n_{e}$ stays constant up to the ionization front, where the gas transitions from ionized to neutral hydrogen. Overall, the electron density ranges from $n_{e} \sim 4-12 \ \rm cm^{-3}$ The temperature profile is smooth distribution, with a central peak near the edge of the photoionized region reaching 22,000 to 24,000~K, which then stays approximately constant at around 14,800~K until reaching the hydrogen ionization front. Note that the inner regions close to the star are simply a cavity of radius 10~pc in the photoionization models, so the precise details of the transition between shock-ionized and photoionized gas are not accurately represented in these models.

The ratio of [\ion{S}{3}] 18 \micron / [\ion{S}{3}] 33 \micron\ also provides an estimate for $n_{e}$ of an HII region. However, this ratio is only sensitive to gas with densities $n_e > 100\  \cm$ \citep{Dudik2007}, thus it does not probe the less dense ionized gas in N76. Our photoionization modeling of N76, however, is able to explore a wider region of parameter space than line ratios alone. \autoref{fig:phy} shows how dramatically the size of the \hii\ region changes for small changes in hydrogen density, illustrating the usefulness of spatially resolved data for characterizing \hii\ regions. 

We also calculate the ionization parameter, $U$, defined as the dimensionless ratio of the hydrogen-ionizing photons to the total hydrogen density. This parameter is often used to characterize \hii\ regions, and is defined as

\begin{equation}
    U = \frac{Q(H)}{4 \pi r_o^{2}n(H) c}\ ,
\end{equation}

\noindent where $r_o$ is the ionized region radius in units of cm, $n(H)$ is the total hydrogen number density in units of \cm , $c$ is the speed of light, and $Q(H)$ is the number of hydrogen-ionizing photons emitted by the central source per second. The ionization parameter we measure for N76 ranges from $U = 1.7 \times 10^{-2}$ to $U = 7.1 \times 10^{-2}$, \changes{which is quite high and comparable to the massive HII region 30 Doradus (see \autoref{fig:line_ratio}). The high radiation fields of AB7 and relatively low density of N76 \newchanges{indicate that N76 has a high ionization parameter, even with a single ionization source.}}

\begin{figure*}
    \centering
    \includegraphics[width=\textwidth]{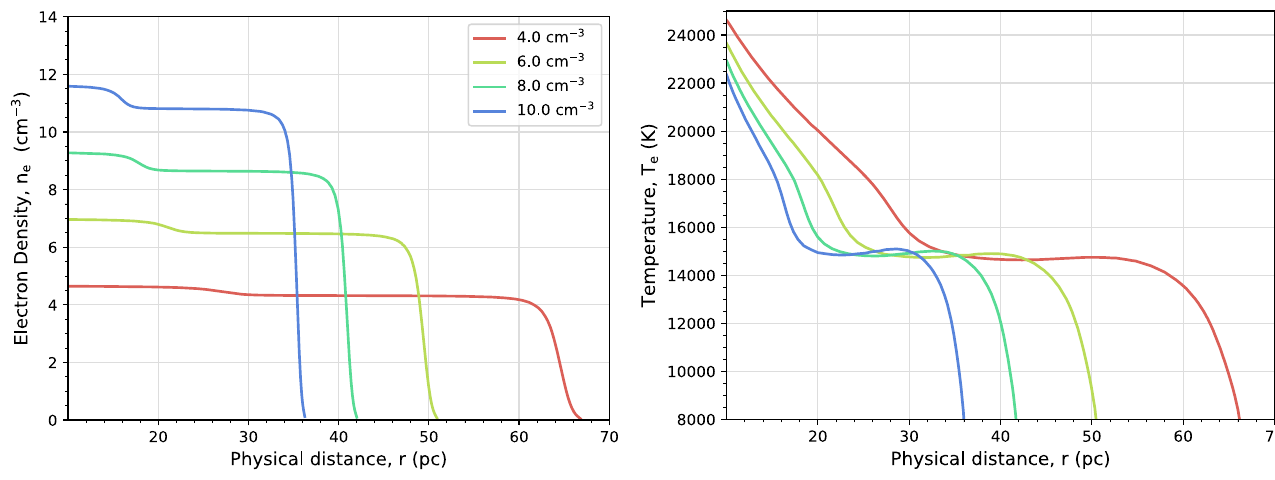}
    \caption{The electron temperature and density for the constant density {\tt Cloudy} models as a function of physical distance away from AB7. Colors follow the same scheme as in \autoref{fig:radprof_den}. }
    \label{fig:phy}
\end{figure*}

\changes{We also note that the {\tt Cloudy} models presented in this work are radiation-bounded, which are defined as \hii\ regions wherein {\tt Cloudy} models end at the ionization front and do not allow appreciable ionizing photons to escape. The {\tt Cloudy} models presented here are an idealized case and represent the average of N76, but it is likely that some parts of the \hii\ region are matter-bounded while others are radiation-bounded. The south-western edge of N76, for example, is much more diffuse than the rest of the nebula, which can allow ionizing photons to escape. A similar morphology is found in N76's \hii\ region neighbor, the supergiant \hii\ region N66 \citep{Geist2022}. }

\subsection{N76 in context}
% \begin{itemize}
%     \item Use the conclusions from this region to speculate on what is implied for other \hii\ regions at low metallicities. 
%     \item Compare with other works (such as \citet[]{Cormier2019} and \citet[]{Hunt2010}. Emphasis the power of resolved observations and the signficance of JWST. 
%     \item Possible \underline{figure} comparing N76 with other low metallicity regions
%     \item make a qualitative description of all the lines -- add a figure of the shell like structure
% \end{itemize}

The N76 nebula is unique for its very luminous central ionizing source, low density ($\sim$10 \cm), and consequent large ($\sim$40~pc) size. In order to compare it to other \hii\ regions, we examine our results through two \changes{different} lenses: observations of the ionized gas in low metallicity dwarf galaxies, and in-depth studies of WR nebulae.

\subsubsection{Comparison with other low metallicity systems}
The largest study focused on understanding the properties of the mid-infrared transitions from gas in low metallicity systems is the Dwarf Galaxy Survey (DGS), which modeled IR observations of 48 nearby galaxies using {\tt Cloudy} \citep{Cormier2015, Cormier2019}. \changes{The results of DGS made considerable progress in the understanding of ionized gas properties in low metallicity systems, but the majority of the sample consists of modeling an unresolved spectrum of each galaxy using a single or two component Cloudy model. While recent, follow up studies use a topological model to infer the statistical properties of the multiphase ISM from the unresolved data \citep{Lebouteiller2022, Ramambason2022, Ramambason2023}, this work instead focuses on the nearby \hii\ region N76 as it is fully resolved and contains a single, well characterized ionization source. The models of N76 presented here provide an ideal case study to understand the impact Wolf-Rayet stars have in low metallicity environments, which are often characterized by active star formation and can contain large number of WR stars \citep[e.g.,][]{Izotov1997}}. 

\changes{We compare the ionized gas properties found in the large surveys of dwarf galaxies below.} \citet{Cormier2019} reports electron densities that range $\rm n_e = 10^{0.5} - 10^{3.0} \ \cm$ and ionization parameters of $\log U = -3.0 \ \rm{to}\  -0.3$ with a correlation such that the highest $U$ values are found in the lowest metallicity galaxies. This trend is also seen in the blue compact dwarf (BCD) galaxies analyzed in \citet{Hunt2010}, where they report elevated [\ion{Ne}{3}]/[\ion{Ne}{2}], [\ion{S}{4}]/[\ion{S}{3}], and [\ion{O}{4}]/[\ion{Si}{2}] ratios when compared to higher metallicity counterparts, indicative of harder radiation fields at low metallicities. Additionally, \citet{Hunt2010} report electron densities ranging from $\rm n_e = 30 - 600 \ \cm$ through the [\ion{S}{3}] 18 \micron / [\ion{S}{3}] 33 \micron\ ratio. These two programs focused on analyzing spatially unresolved infrared spectroscopy of galaxy-wide measurements in order to form conclusions about the global ionized gas properties in dwarf galaxies. With a metallicity of 12 + log(O/H) $\sim$ 8.0, N76, as part of the SMC, falls at the median of the metallicity ranges investigated in these studies. \changes{The resolved measurements of N76} are well in the range of the globally integrated properties determined by these surveys, although the electron density in N76 is at the lower boundary found in these studies. \newchanges{This suggests that the brightest \hii\ regions can dominate the excitation properties of the ISM in the ionized gas of these galaxies. } 

We also compare N76 to photoionization modeling studies of nearby dwarf galaxies that are more spatially resolved than those just described. To be consistent, all of these studies are done using similar IR spectroscopic observations and modeled with the {\tt Cloudy} photoionization code. \citet{Indebetouw2009} model 30~Doradus in the Large Magellanic Cloud (LMC), the largest supergiant \hii\ region in the nearby universe. They find a relatively high electron density, $n_{e} = 10^{2.4} - 10^{2.7} \ \cm$ and a hard radiation field with $\log U = -1.7 \ \rm{to} -1.3$. Similarly, \citet{Polles2019} investigates the ionized gas in five individual star-forming clumps in IC10, and their {\tt Cloudy} modeling indicates $n_{e} = 10^{2.0} - 10^{2.6} \ \cm$ and $\log U = -3.8 \ \rm{to} -1.0$. Lastly, \citet{Dimaratos2015} focus their study on NGC 4214 and identify a separate central and southern star forming region, calculating lower densities of $n_{e} = 440$ and $n_{e} = 170$, respectively, and ionization parameters of $U = -2.3$ and \changes{$U = -2.7$}. Taken as a whole, N76 has a lower electron density than found in those objects, \newchanges{with somewhat larger} ionization parameters. 

In order to summarize the infrared {\tt Cloudy} modeling results across these different surveys and studies, we plot the [\ion{Ne}{3}] 15 \micron / [\ion{Ne}{2}] 12 \micron\ and the {\tt Cloudy} model ionization parameter, $U$, in \autoref{fig:line_ratio}. This ratio traces the intensity of the radiation field and, in principle for a single main sequence star hardness should correlate well with $U$. We also plot a summary of the [\ion{Ne}{3}] 15 \micron / [\ion{Ne}{2}] 12 \micron\ ratio from a sample of high metallicity galaxies \citep{Dale2009, Inami2013}. While these analyses did not produce an ionization parameter, we do show their range in [\ion{Ne}{3}] 15 \micron / [\ion{Ne}{2}] 12 \micron\ line ratio. We confirm that the higher metallicity galaxies have a much lower [\ion{Ne}{3}] 15 \micron / [\ion{Ne}{2}] 12 \micron\ line ratio than the low metallicity systems. N76 and 30~Doradus are at the upper third of the observed line ratio, with [\ion{Ne}{3}] 15 \micron / [\ion{Ne}{2}] 12 \micron\ $\sim$ 10-15, indicating hard radiation fields in these \hii\ regions. \changes{Many of the galaxies in the DGS sample have a similar ionization parameter and line ratio as these highly energetic HII regions, suggesting that conditions in these galaxies may be very similar to the resolved studies of N76 and 30Dor.}

When examining scales of $\sim$200~pc, \citet{Cormier2019} find evidence of additional, low ionization components of ionized gas from their {\tt Cloudy} models. Similarly, \citet{Polles2019} and \citet{Dimaratos2015} find excess emission of [\ion{Ne}{2}] that they attribute to a diffuse ionized gas component. In N76, our {\tt Cloudy} models also underpredict [\ion{Ne}{2}], providing more evidence for an additional low ionization source that fuels the diffuse ionized gas in these galaxies.

\begin{figure}
    \centering
    \includegraphics[width=\columnwidth]{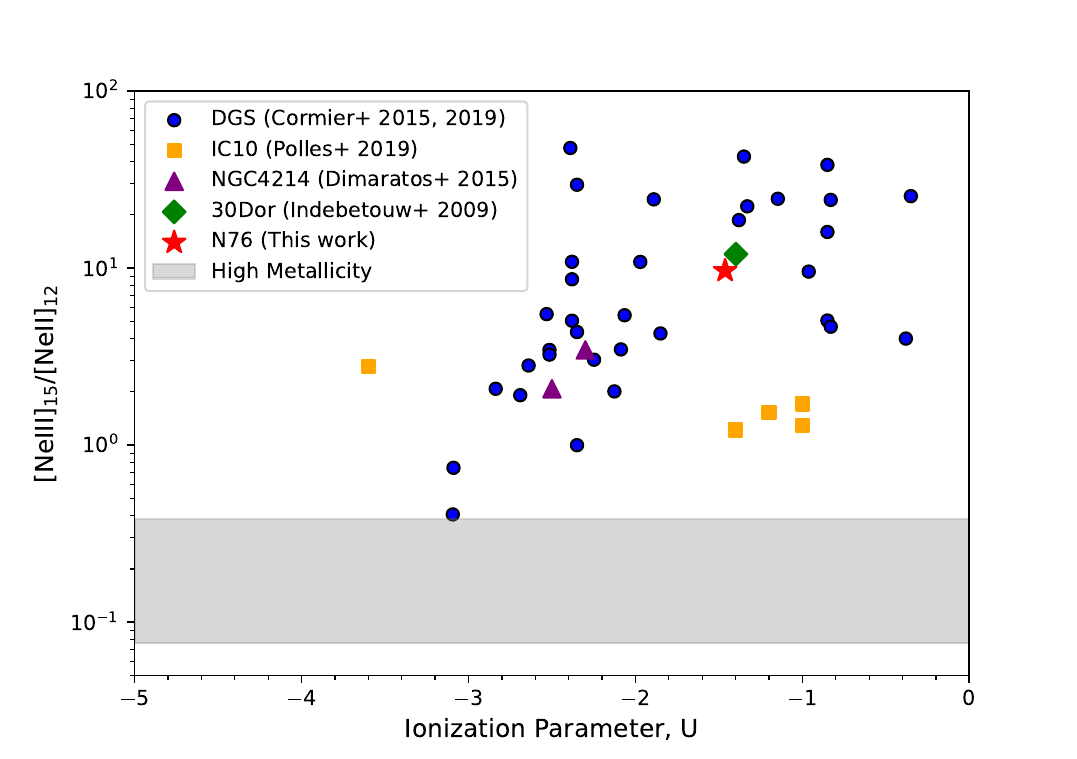}
    \caption{Comparing the [\ion{Ne}{3}] 15 \micron / [\ion{Ne}{2}] 12 \micron\ ratio to the ionization parameter of a variety of low metallicity systems, including the Dwarf Galaxy Survey \citep{Cormier2015, Cormier2019}, IC10 \citep{Polles2019}, NGC 4214 \citep{Dimaratos2015}, 30 Doradus \citep{Indebetouw2009}, and N76 (this work). These studies include a mixture of integrated data from entire galaxies (DGS), to individual measurements in divided regions of these galaxies (IC10 and NGC 4214), to single HII regions (N76 and 30~Dor). Overall, the [\ion{Ne}{3}] 15 \micron / [\ion{Ne}{2}] 12 \micron\ ratio shows that the low metallicity systems plotted have much harder radiation fields than the averages shown for the higher metallicity galaxies from \citet{Dale2009} and \citet{Inami2013}.}
    \label{fig:line_ratio}
\end{figure}

% \autoref{fig:line_ratio} shows the 15 \micron / [\ion{Ne}{2}] 12 \micron\ line ratio of N76 and a variety of other low metallicity systems. One of the goals of this study was to model the low metallicity ionized gas of a single  HII region as compared to aggregate modeling that creates an ionized gas model from an integrated galaxy spectrum. The 15 \micron / [\ion{Ne}{2}] 12 \micron\ line ratio traces the ionization parameter defined by cloudy. Each of the comparison observations were taken to be low metallicity systems that ranged in model scale. N76 follows the same trends very closely. 

\subsubsection{Comparison with other Wolf-Rayet nebulae}
Most well-studied WR nebula are defined as bubble nebulae, characterized by thin ionized shell covering a large cavity, excavated by the strong stellar winds of the central WR star. Optical IFU measurements of the Milky Way WR nebula NGC 6888 show a three-shell structure centered around a 400~pc evacuated cavity: an elliptical inner broken shell formed by shocks from interactions between the WR and supergiant winds, an outer spherical shell that represents the wind-blown bubble, and a skin of ISM material surrounding the nebula \citep{Fernandez2012}. Through photoionization modeling, \citet{Fernandez2012} calculate an electron density of $n_e = 400 \ \cm$ for the outer wind-blown bubble shell. Infrared spectroscopy described in \citet{Rubio2020} characterize NGC 6888 further, corroborating the density found for the outer shell in \citet{Fernandez2012} and identifying the density of the ISM skin as $n_e = 180 \ \cm$. 

The overall morphological picture of NGC 6888 is not dissimilar to that of N76: a wind-shocked inner cavity caused by the powerful stellar winds of the WR star that transitions into a photoionized gas region and the surrounding ISM. However, N76 has a much smaller cavity (10~pc in our models compared to 400~pc), a lower density photoionized region with $n_e = 4-10 \ \cm$, and this photoionized shell is rather thick compared to the cavity size (35~pc). These differences may be related to the age of N76. \citet{Shenar2016} estimates that AB7 is 3.4~Myr old, 1.3 Myr younger than WR 136 that powers NGC 6888 \citep{Moore2000}. The size of a stellar wind-blown bubble directly correlates with its age \citep{Weaver1977, Freyer2003, Freyer2006}, suggesting that the younger N76 nebula will be smaller than NGC 6888. Observations of WR nebulae in the LMC show the traditional bubble structure, i.e. a large cavity with a thin, photoionized shell of material in the majority of nebulae \citep{Hung2021}. 

%could be due in part to the lower metallicity of the ISM in the SMC; the lower dust abundances may allow photons to penetrate deeper, forming a larger nebula, similar to how low metallicity PDRs are larger than their higher metallicity counterparts \citep[e.g.,][]{Poglitsch1995}.  

Since the regions around WR nebulae contain material blown off by the powerful stellar winds of the star, their abundances can be enhanced, yielding elevated nitrogen abundances in those with a central WN star \citep[e.g.,][]{Fernandez2012, Stock2014}. AB7 contains a WN4 type star, but we do not have a nitrogen spectral line to probe the nitrogen abundance. The abundances described in \autoref{tab:abun} do a satisfactory job of predicting the emission line surface brightness so we do not have evidence of elemental enhancement in N76. Optical spectroscopy of N76 by \citet{Naze2003}, however, show a 40\% elevated N/O ratio compared to global SMC values, suggesting that the chemical make up of N76 is influenced by AB7.

\section{Summary and Conclusions}
\label{sec:conclusions}
We use {\em Spitzer} IRS spectroscopic maps of mid-infrared transitions and {\tt Cloudy} to model the photoionization in the \hii\ region N76 in the Small Magellanic Cloud. The ionization source is AB7, a binary composed of a WN4 nitrogen-rich Wolf-Rayet and an O6\,I supergiant star, and we use PoWR models \citep{Todt2015} to estimate its SED and luminosity \citep{Shenar2016}. AB7 contains one of the hottest WR stars in the SMC. We assume the N76 nebula is spherically shaped and project the predicted {\tt Cloudy} line surface brightness as a function of distance to the source on the plane of the sky. We only consider emission from the ionized gas. The results from the photoionization models match the spatially resolved images of most ionized emission lines measured from \spitzer\ IRS and \herschel\ PACS observations, with just density as the free parameter. Our conclusions are as follows:

\begin{enumerate}
\itemsep-0.5em
    \item Comparing the models to the observed spatial distribution of the measured emission line brightness allows us to determine much more precise densities than just using the total line intensity (\autoref{fig:radprof_den} and \autoref{tab:lines}).  
    \item Constant density {\tt Cloudy} photoionization models reproduce most of the ionized gas emission line intensity distributions ([\ion{S}{3}], [\ion{Ne}{3}], [\ion{S}{4}], [\ion{Ne}{5}], [\ion{O}{4}], and [\ion{O}{3}]) for a narrow hydrogen density between $\rm n_{H} = 4 \ cm^{-3} - 10 \ cm^{-3}$ (\autoref{fig:linemaps}).

    \item The {\tt Cloudy} photoionization models require a 10~pc cavity of hot, shocked plasma originating from a stellar wind-blown nebula in order to predict the spatial distribution of the ionized gas emission lines \citep{Weaver1977, Freyer2003, Freyer2006}. 

    \item The high ionization lines, [\ion{O}{4}] and [\ion{Ne}{5}] ($\rm E_{ion} > 50 \ eV$), are well predicted through the Cloudy models using the SED modeled in \citet{Shenar2016}. The flux of these lines is highly sensitive to changes in the high energy tail of the WR SED. 
    
    % \item The intensities of the high ionization lines, [\ion{O}{4}] and [\ion{Ne}{5}], can be explained by just photoionization. We postulate that these ions are produced by shock ionization in the wind-blown portion of the WR nebula \citep[see also][]{Weaver1977, Freyer2003, Freyer2006} or a harder ionizing SED for the central source \citep[e.g.,][]{Crowther1999, Schaerer1998, Schaerer1999}. To explain the observations requires a 15~pc in radius wind shocked cavity in the center of N76. 
    
    \item The neutral-dominated emission lines (lines coming from ionic species with an ionization potential less than hydrogen, 13.6 eV), can originate from both the ionized gas and the neutral gas. Our modeling finds that very little of the observed [\ion{C}{2}] emission comes from the ionized gas ($<5\%$), while most (91\%) of the [\ion{Si}{2}] emission can be produced \changes{in the transition from \hii\ region to PDR.} 

    \item The [\ion{Ne}{2}] line is under-predicted in our {\tt Cloudy} models, suggesting the excess emission may arise from the diffuse ionized gas not associated with N76 \citep[e.g.,][]{Dimaratos2015, Cormier2019, Polles2019}.
    
    \item We observe a \changes{feature at $25.25$~\micron\ that we tentatively identify} as [\ion{S}{1}] in the high ionization zone of the N76 nebula (see \autoref{fig:linemaps}). We explore possible causes for [\ion{S}{1}] and alternative possibilities for the line in \S\ref{sec:SI_ideas}, including the possibility of producing neutral sulphur from dust destruction. None of these satisfactorily explains the line. 
    \newchanges{\item The physical conditions of N76 derived from the {\tt Cloudy} models characterize a hot \hii\ region with a maximum electron temperature range from $T_e \sim 22,000-24,000 \rm K$, a relatively small density range of $n_{e} \sim 4-12 \ \rm cm^{-3}$, and high ionization parameters of $log(U) = -1.15 - -1.77$ (\S\ref{sec:phy_con}).}
    
    \item Low metallicity galaxies have harder radiation fields than their higher metallicity counterparts. \changes{The N76 region has a similar ionization parameter to samples of low metallicity galaxies, suggesting that the bright \hii\ regions dominate the spectrum of individual galaxies (see \autoref{fig:line_ratio})}.  
\end{enumerate}

Our work shows that just photoionization from low metallicity WR stars can produce appreciable [\ion{O}{4}] and [\ion{Ne}{5}] emission in the surrounding ISM. Thus N76 is a vital case study in determining the role of WRs in producing large fluxes of high ionization ions ($\rm E_{ion} > 50 \ ev$) found in low metallicity Extreme Emission Line Galaxies (EELG) and blue compact dwarf galaxies \citep[e.g.,][]{Thuan2005,Izotov2012, Izotov2021, Kehrig2015, Kehrig2018, Leitherer2018, Olivier2022}. Due to the rarity of WRs in population synthesis modeling \citep[e.g.,][]{Leitherer2014}, WR stars are often ruled out as a contributor to the high ionization lines. Our work suggests that the choice of WR model can make a significant impact on the high energy tail of the SED and consequently the amount of [\ion{Ne}{5}] and [\ion{O}{4}] produced, suggesting that special care should be taken when modeling WR stars in unresolved systems. 

\begin{acknowledgements}
We would like to thank T. Shenar for providing the SED of AB7 from \citet{Shenar2016}. \changes{We would also like to thank the anonymous referee for providing feedback that greatly improved the manuscript. }

ET and ADB acknowledge support from NSF-AST2108140 and NASA ADAP 80NSSC19K1015.
RI acknowledges support from NSF-AST2009624.
M.R. wishes to acknowledge support from ANID(CHILE) through FONDECYT grant No1190684 and partial support from ANIDA Basal FB210003.

\end{acknowledgements}

\software{astropy (The Astropy Collaboration 2013, 2018, 2022), CUBISM (v1.8; Smith et al. 2007a), TinyTim (Krist et al. 2011), {\tt Cloudy} (Ferland et al. 2013), PAHFIT (Smith et al. 2007b)}

\bibliography{references}{}
\bibliographystyle{aasjournal}

\appendix
\section{PSF Matching Procedure}
\label{sec:app-psf}

We follow the procedure described by \citet{aniano2011} to create custom convolution kernels that will transform an image with a narrower PSF to a broader PSF.

Mathematically, we generate a kernel, $K$, that when convolved transforms $\Psi_{B}$, the narrower PSF, to $\Psi_{A}$, the broader PSF:

% Each line is matched to the PSF of the [\ion{Si}{2}] 34.8 \micron\ line, which has the broadest PSF out of the spectral lines detected with \spitzer\ IRS. To summarize, we generate a kernel, $K$, that when convolved transforms $\Psi_{B}$, the PSF of any wavelength line lower than 34 \micron , to $\Psi_{\rm{[\ion{Si}{2}]}}$, the PSF of the [\ion{Si}{2}] line:

\begin{equation}
\label{eq:kernel}
    \Psi_{A} = \Psi_{B} \ast K(B \rightarrow A).
\end{equation}

To calculate the convolution kernel, $K$, we take the Fourier Transform (FT) and solve: 

\begin{equation}
\label{eq:psf_match}
    K(B \rightarrow A) = \rm{FT^{-1}} \bigg(\rm{FT}(\Psi_{B}) \times \frac{1}{\rm{FT}(\Psi_{A})}\bigg).
\end{equation}

Computing this kernel numerically while ensuring its accuracy and stability requires several steps outlined in \citet{aniano2011} and are summarized below for completeness. 

\subsubsection{Preparing the PSFs}
The IRS PSFs are from version 2.0 of the \spitzer\ Tiny Tim software\footnote{\url{https://irsa.ipac.caltech.edu/data/SPITZER/docs/dataanalysistools/tools/contributed/general/stinytim/}}.  We create model PSFs for the full wavelength range of IRS, 5~\micron\ - 38~\micron . We note that these are simulated PSFs and there are some differences between the predicted PSF for IRS and the measured PSF from calibration stars, but these differences are negligible for our application \citep{Pereira-Santaella2010}. The PSFs range in pixel scale and image size, depending on the module. In order to calculate the convolution kernel, $K$, we regrid all PSFs to a common pixel scale of $0.2 \arcsec$ and an image size of $1859 \ \times \ 1859$ pixels. The PSF matching algorithm outlined in \citet{aniano2011} works best when the PSFs are rotationally symmetric. We circularize the PSFs by rotating the PSF 14 times and averaging after each rotation in order to make the final PSF invariant for any angle that is a multiple of $360^{o}/2^{14} = 0.022^{o}$.  

\subsubsection{Creating the convolution kernel}
We compute the Fourier Transform (FT) of the narrower PSF, $\Psi_B$, and the broader PSF, $\Psi_{A}$. We use the \texttt{PYTHON} package \texttt{FFTPACK} for all FT calculations. The FT of the target PSF is in the denominator (see \autoref{eq:psf_match}) and will amplify any small high-frequency components of the FT when creating the convolution kernel, $K$. We account for this by introducing a low-pass filter, $f$, in kernel construction:

\begin{equation}
    f(k) = 
    \begin{cases}
    1 & \mathrm{for} \ k \leq k_{L} \\
    \frac{1}{2} \times \bigg[1 + \cos\big( \pi \times \frac{k-k_{L}}{k_{H} - k_{L}}\big)\bigg] & \mathrm{for} \  k_{L}  \leq k \leq k_{H} \\ 
    0 & \mathrm{for} \ k_{H} \leq k
    \end{cases}
\end{equation}
where k is the spatial frequency in the Fourier domain, $k_H$ is the high frequency cutoff for the filter, and $k_L = 0.7 \times k_H$ is the low-frequency cutoff. We create the convolution kernel by modifying \autoref{eq:psf_match} with the low-pass filter:

\begin{equation}
\label{eq:psf_match}
    K(B \rightarrow A) = \rm{{FT^{-1}} \bigg(\rm{FT}(\Psi_{B}) \times \frac{1}{\rm{FT}(\Psi_{A})}} \times f(k) \bigg).
\end{equation}
repeating the process by looping over the narrower PSFs for $\Psi_B$. Lastly, we normalize the kernel to unity to ensure flux conservation.

\subsubsection{Testing the kernel}
We test the convolution kernel for accuracy and stability through the metrics defined by \citet{aniano2011}. The convolution of the narrower PSF, $\Psi_B$, and the kernel, $K(B \rightarrow A)$, should reproduce the broader PSF $\Psi_{A}$ (see \autoref{eq:kernel}). Further, kernels with many negative values, which redistribute flux, can be unstable. \citet{aniano2011} define two parameters, $D$ (their equation 20), the integral of the difference between the convolved PSF and the target PSF, and $W^{-}$, a sum of the negative values in the kernel, to evaluate the kernels. We calculate these metrics and find very low values for $D$ ($<$ 0.01 for all kernels) and satisfactory values for $W^{-}$ (ranges between 0.2 - 1.0). \citet{aniano2011} report that $W^{-} < 1.2$ are stable. Further, the value chosen for the filtering function $k_H$ affects these metrics, where an increase in $k$ leads to a lower $D$ but a higher $W^{-}$. When constructing kernels for each wavelength, we use a range of $k_{H}$ values and select a $k_{H}$ with the minimum of both $W^{-}$ and $D$. The $k_{H}$ filtering value ranges from $0.7 - 1.0$ for the final convolution kernels.  

\subsubsection{The final PSF matched images}
We produce two sets of PSF matched line images: one that provides maximum uniform resolution by matching to the PSF of the [\ion{Si}{2}] transition (our longest wavelength line), and one that matches everything to a 12\arcsec\ Gaussian for ease of comparison with other data. In our analysis we use the 12\arcsec\ Gaussian PSF. The [\ion{Si}{2}] 34.8 \micron\ line has the broadest PSF out of the spectral lines detected with \spitzer\ IRS with a FWHM of 7.6\arcsec. We match each narrower IRS PSF to the PSF of the [\ion{Si}{2}] line. Then, we create and apply a convolution kernel to go from the [\ion{Si}{2}] PSF to a 2D Gaussian PSF. We find that a 2D Gaussian with a FWHM of 12\arcsec\ best matches the [\ion{Si}{2}] PSF while balancing kernel stability. 

Each emission line image is convolved with its custom kernel using the \texttt{ASTROPY} package \texttt{CONVOLUTION}. We then regrid each image to the coordinates of the [\ion{Si}{2}] image to match its pixel scale and to simplify the analysis. For the analysis we also mask non-physical bright artifacts and the emission from the SNR E0102-72. We use 12\arcsec\ Gaussian PSF-matched final images for the analysis in this work. Note that \autoref{fig:linemaps} presents the original resolution images to show the maximum detail in the morphology of N76. 

In addition to creating the PSF matched line maps that are described in $\S$\ref{sec:linemaps}, we also produce a PSF matched cube for the full IRS spectrum, including all orders: SL1, SL2, LL2, LL1. We follow the same procedure on each frame of the original data cubes. A spectrum of the central position of the N76 nebula from the PSF-matched cube is presented in \autoref{fig:spec}.

We PSF match the IRS \spitzer\ data, but not the \herschel\ PACS images (see $\S$\ref{sec:PACS}). The emission line images from PACS have larger PSFs that would degrade too much the resolution of the IRS maps if we were to match each image to the largest \herschel\ PSF.

\section{Effect of asymmetries in N76}
\label{sec:app}

In order to compare the {\tt Cloudy} models directly to various emission lines in N76, we need to assume spherical symmetry. However, it is clear that the images of the emission lines show spatial asymmetries in 
\autoref{fig:linemaps}. The eastern side of N76 is brighter in almost every IR line and the western edge is less well-defined, with more diffuse, extended features. To explore how these asymmetries affect the {\tt Cloudy} modeling, we show here the results for a narrow strip that intersects with the AB7 and designate it the central point. We then sample the surface brightness to the left (eastern edge) and right (western edge) of the strip and report it in \autoref{fig:strip}.

The asymmetric profiles show that the western edge prefers a lower density model in almost all cases (except for [\ion{O}{1}], [\ion{C}{2}], and [\ion{Ne}{2}]). The idealized symmetric model presented in the radial profiles in \autoref{fig:radprof_den} is simply an average of these two sides. The difference between these densities is relatively low, however, with the higher density side preferring a model only $4 \ \cm$ larger. This is typically a $\pm20-30\%$ change in the result of the spherically symmetric modeling. 

\label{fig:strip}
\begin{figure*}[b]
    \centering
    \includegraphics[width=0.65\textwidth]{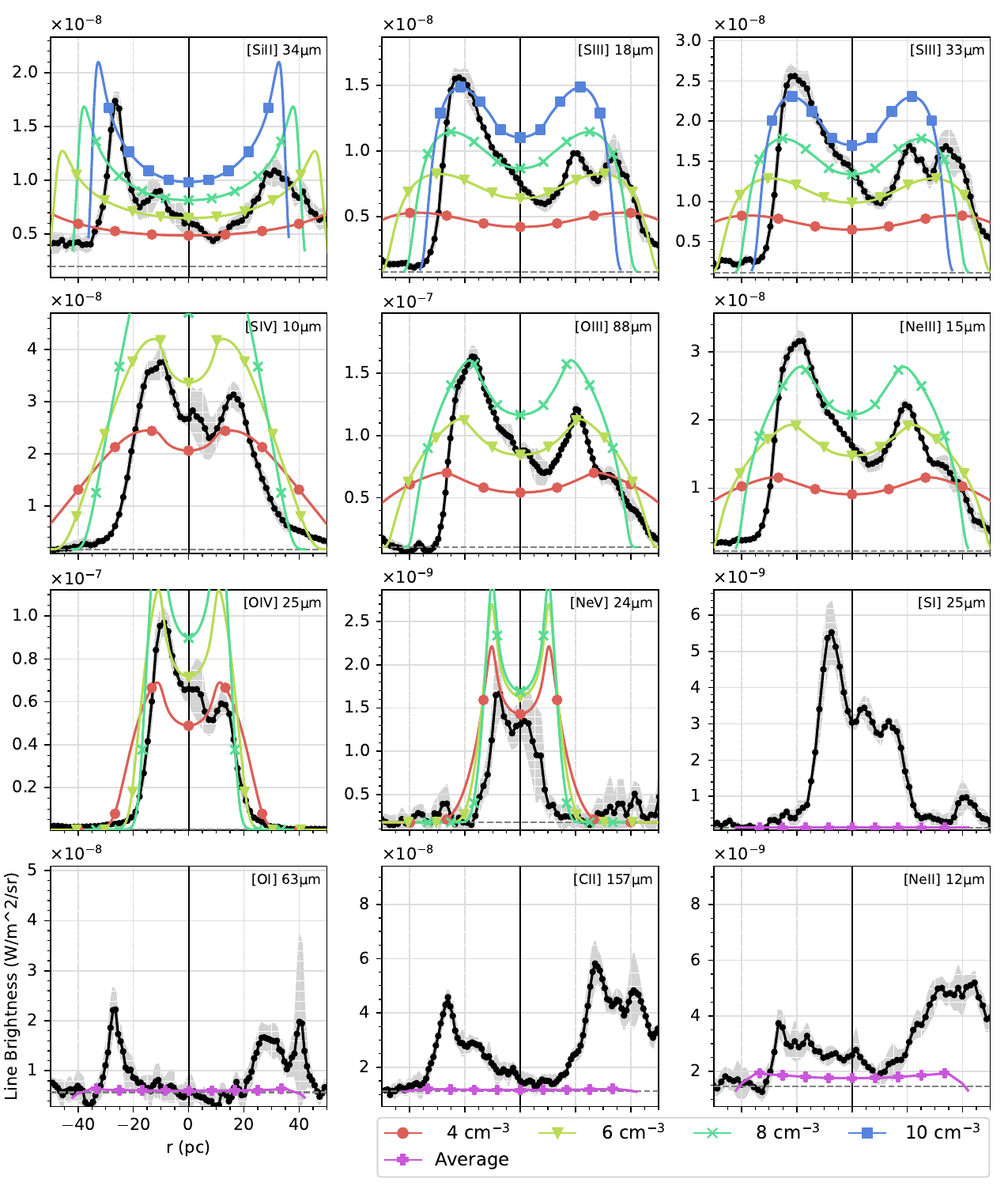}
    \caption{The surface brightness distribution of emission lines taken as a strip across the nebula with AB7 in the center and the eastern and western edges on the left and right hand of the plot, respectively. This shows the asymmetry of N76, where the eastern edge is brighter than the western side. The {\tt Cloudy} models prefer a slightly lower density for the western side.}
    \label{fig:strip}
\end{figure*}

%% This command is needed to show the entire author+affiliation list when
%% the collaboration and author truncation commands are used.  It has to
%% go at the end of the manuscript.
%\allauthors

%% Include this line if you are using the \added, \replaced, \deleted
%% commands to see a summary list of all changes at the end of the article.
%\listofchanges

\end{document}